\documentclass[12pt,preprint]{aastex}
\received{2004 December 29}
\begin{document}

\title{Lithium and Lithium Depletion \\
in Halo Stars on Extreme Orbits}

\author{Ann Merchant Boesgaard\altaffilmark{1}} 
\affil{Institute for Astronomy, University of Hawai`i at M\-anoa, \\ 
2680 Woodlawn Drive, Honolulu, HI {\ \ }96822 \\ } 
\email{boes@ifa.hawaii.edu}

\author{Alex Stephens\altaffilmark{1,}\altaffilmark{2}}
\affil{Institute for Astronomy, University of Hawai`i at M\-anoa, \\ 
2680 Woodlawn Drive, Honolulu, HI {\ \ }96822 \\ } 
\email{Alex\_C\_Stephens@yahoo.com}

\and

\author{Constantine P. Deliyannis\altaffilmark{1}}
\affil{Department of Astronomy, Indiana University\\
727 East 3rd Street, Swain Hall West 319, Bloomington, IN {\ \ }47405-7105}
\email{con@athena.astro.indiana.edu}

\altaffiltext{1}{Visiting Astronomer, W.~M.~Keck Observatory, jointly operated
by the California Institute of Technology and the University of California}
\altaffiltext{2}{Visiting Astronomer, Kitt Peak National Observatory, National
Optical Astronomy Observatory, operated by the Association for Research in
Astronomy, Inc., under contract with the National Science Foundation.}

\begin{abstract}
We have determined Li abundances in 55 dwarfs and subgiants that are
metal-poor ($-$3.6 $<$ [Fe/H] $<$ $-$0.7) and have extreme orbital kinematics.
Our purpose is to examine the Li abundance in the Li-plateau stars and its
decrease in low-temperature, low-mass stars.  For the stars in our sample we
have determined chemical profiles in Stephens \& Boesgaard (2002).  The Li
observations are primarily from the echelle spectrograph on the 10 m Keck I
telescope with HIRES covering 4700 - 6800 \AA\ with a spectral resolution of
$\sim$48,000.  The spectra have high signal-to-noise per pixel from 70 to 700,
with a median of 140.  The Li I resonance doublet was detected in 42 of the 55
stars.  Temperatures were found spectroscopically by Stephens \& Boesgaard
(2002).  Abundances or upper limits were determined for all stars with typical
errors of 0.06 dex.  Corrections for the deviations from non-local
thermodynamical equilibrium for Li in the stellar atmospheres have been made
which range from $-$0.04 to +0.11 dex.  Our 14 dwarf and turn-off stars on the
Li plateau with temperatures greater than 5700 K and [Fe/H] $<$$-$1.5 give
A(Li) = log N(Li)/N(H) + 12.00 of 2.215 $\pm$0.110, consistent with earlier
results.  We find a dependence of the Li abundance on metallicity as measured
by [Fe/H] {\it and} the Fe-peak elements Cr and Ni, with a slope of
$\sim$0.18.  We have examined the possible trends of A(Li) with the chemical
abundances of other elements and find similar dependences of A(Li) with the
alpha elements, Mg, Ca, and Ti.  These slopes are slightly steeper at
$\sim$0.20, resulting from an excess in [$\alpha$/Fe] with decreasing [Fe/H].
For the n-capture, rare-earth element, Ba, we find a relation between A(Li)
and [Ba/H] which has a shallower slope of $\sim$0.13; over a range of 2.6 dex
in [Ba/H], the Li abundance spans only a factor of two.  We have also examined
the possible trends of A(Li) with the characteristics of the orbits of our
halo stars.  We find no trends in A(Li) with kinematic or dynamic properties.
For the stars with temperatures below the Li plateau there are several
interesting results.  The group of metal-poor stars possess, on average, more
Li at a given temperature than metal-rich stars.  When we divide the cool
stars into smaller subsets with similar metallicities, we find trends of A(Li)
with temperature for the different metallicity groups.  The decrease in A(Li)
sets in at hotter temperatures for the higher metallicity stars than for the
lower metallicity stars.  The increased Li depletion in cooler stars could be a
result of the increased action of convection since cooler stars have deeper
convection zones.  This would also make it easier for additional mixing
mechanisms, such as those induced by rotation, to have a greater effect in
cooler stars.  Since the model depth of the convection zone is almost
independent of metallicity at a given effective temperature, the apparent
metallicity-dependence of the Li depletion in our data may be pointing to
subtle but poorly understood mixing effects in low mass halo dwarfs.
Predictions for Li depletion from standard and non-standard models seem to
underestimate the degree of depletion inferred from the observations of the
cool stars.
\end{abstract}


\section{Introduction}

The study of Li in metal-poor stars has produced a cascade of papers after the
initial studies by the Spites (Spite \& Spite 1982; Spite, Maillard \& Spite
1984).  Most of the papers have been concerned primarily with the
determination of the value the amount of Li produced during the Big Bang.
Some recent examples include Ryan et al.~(1996); Bonifacio \& Molaro (1997);
Ryan \& Deliyannis (1998); Ryan et al.~(1999); Ryan et al.~(2000); Ryan
(2002); Zhang \& Zhao (2003); Mel\'endez \& Ram\'\i rez (2004).  This
primordial Li, called A(Li$_p$), where A(Li) = log N(Li)/N(H) +12.00, is of
importance in our understanding of the conditions during the Big Bang and
subsequently.  Recently Coc et al.~(2004) have compared the predictions from
the {\it Wilkinson Microwave Anisotropy Probe}, ({\it WMAP}; Spergel et
al.~2003) with the observations of A(Li$_p$) from Ryan et al.~(2000) (and
Ryan, Norris \& Beers 1999) which reveal a significant discrepancy.  Coc et
al.~(2004) find {\it WMAP}+SBBN (standard Big Bang nucleosynthesis) produces a
value for A(Li$_p$) of 2.62 $\pm$0.05 compared to the observed value of 2.0 -
2.2.  Mel\'endez and Ram\'\i rez (2004) suggest a marginal consistency with
the observed A(Li$_p$) of 2.37 through their use of a hotter temperature
scale.

The 55 stars selected for this Li project are all metal-poor ($-$3.6 $<$
[Fe/H] $<$$-$0.7) dwarfs and subgiants with extreme orbital characteristics.
The orbits have one or more of these criteria: extreme retrograde velocities,
from the outer Galactic halo, high altitudes above the Galactic plane.  In
addition there are a few stars with intermediate characteristics for
comparison.  The stars are presently in the solar neighborhood, but they are
transients from remote parts of the Galaxy.

This research is motivated by several new aspects of Li in metal-poor stars.
1) Our sample includes a large subset of cool stars so we can investigate the
decline in Li in low-temperature, low-mass stars.  2) Because our selection of
stars are halo stars with extreme orbits, we can examine if there are
different Li abundances related to orbital dynamics.  3) Abundances of ten
elements have been determined so we have a sample with an unusually complete
chemical profile, including $\alpha$-fusion products, Fe-peak elements, and
neutron-capture elements (Stephens \& Boesgaard 2002, hereafter SB02); thus we
can look for trends of elemental abundances with Li.  4) Some of the abundance
findings in SB02 can be attributed to chemical-enrichment products from SNII
so we can check if there is any Li-enrichment from the $\nu$-process (Timmes
et al. 1995).

\section{Observations and Data Analysis}

The spectra for this research were obtained primarily from the W.~M.~Keck 10 m
telescope with the high-resolution spectrometer, HIRES (Vogt et al. 1994) and
have a spectral resolution of $R \sim$ 48,000 and a wavelength coverage of
4500 - 6800 \AA.  Three of the stars were observed with the KPNO Mayall 4 m
telescope and the echelle spectrometer, ECH (Pilachowski \& Willmarth 1981)
with a resolution of $R \sim$ 35,000.  The observations were made at high
ratios of signal-to-noise pixel$^{-1}$ (S/N): between 70 and 700 in the order
containing the Li I resonance line at $\lambda$6707.  The median S/N is 140
and 67\% of the stars have S/N between 100 and 200 with another 20\% above
200.  The details of the observations and data reduction procedures can be
found in SB02.

Examples of some of the spectra in the Li region are shown in Figure 1.  These
spectra are of stars that are hotter than 5700 K, i.e. stars on the Li
plateau.  The stars shown have a range in metallicity of an order of
magnitude; the spectra shown are of stars from the outer halo and/or from high
in the halo (see figure caption).

\section{Abundances}

The spectrum of each program star was examined for Li I resonance line.  The
6707 \AA\ blend was confidently detected in 42 of the 55 stars while upper
limits were calculated for the remaining 13.  The Cayrel (1988) formula (as
recast in Deliyannis et al.~1993) was used to calculate upper limits given the
S/N in the Li order of the spectrum.  The Li equivalent widths were measured
in IRAF routine {\it splot} assuming a gaussian profile.  Table 1 lists each
star, the S/N in the Li order, the model parameters used, the measured Li
equivalent width or the 3$\sigma$ upper limit along with the measurement
error, the calculated Li abundance or upper limit, A(Li), the NLTE abundance,
A(Li)$_{NLTE}$ (see below), and the 1$\sigma$ error in A(Li).

The grid of Kurucz model atmospheres was used to produce interpolated models
with the stellar parameters in Table 1.  These parameters are those derived in
SB02.  Temperatures were determined spectroscopically from some 30-40 weak Fe
I lines of a range of excitation potentials.  Gravities were found by forcing
neutral and ionized lines of both Fe and Ti to give the same elemental
abundances.  In SB02 our temperatures were compared with those of Carney et
al.~(1994) and Alonso et al.~(1996) and found fairly good agreement.  The
stars with the largest differences were those with $E(B-V) \geq$ 0.05, but the
spectroscopic temperatures are not affected by large and uncertain reddening
corrections as are the photometric temperatures.  There is a slight systematic
trend with the Carney et al.~(1994) temperatures such that our temperatures
are higher by $\sim$100 K at 5000 K, by $\sim$50 K at 5500 K, but in excellent
agreement at 6000 K.  In the comparison with Alonso et al.~(1996) the trends
with temperature are such that our temperatures are hotter by $\sim$40 K at
5000 K, in excellent agreement at 5500 K and $\sim$ 50 K cooler at 6000 K.
These differences seem minor as none of the scales is more accurate than
$\sim$100 K.  We have made comparisons with other temperature determinations
by Fulbright (2000), Bonifacio \& Molaro (1997), Ryan et al.~(1999), and
Akerman et al.~(2004) for the handful of stars in common with each of those
samples.  The various methods and calibrations used by different researchers
show a spread in derived temperatures of as much as 100 - 200 K.  We proceed
here with our spectroscopically derived temperatures because they have been
done consistently and are insensitive to reddening.  Only weak Fe I lines, log
W$_{\lambda}$/${\lambda}$ less than $-$5.15 (or W$_{\lambda}$ $<$ 40 m\AA\
near 5600 \AA), were used to derive $T_{\rm eff}$ so the same set of Fe I
lines could not be used in the metal-poor stars as in the metal-rich stars.
However, the line sets had many lines in common going along the metallicity
ladder.  The total number of Fe I lines in the line set was 133.

Abundances were determined from the Li equivalent widths through use of MOOG
(Sneden 1973, http://verdi.as.utexas.edu/moog.html).  The Li abundances are
quite insensitive to uncertainties in the derived values of log g, [Fe/H], and
microturbulent velocity, $\xi$.  The effect of decreasing log g by $-$0.5 dex
is at most a decrease of 0.01 in A(Li), but the uncertainty in log g averages
0.33 $\pm$0.07 dex (not as much as 0.5 dex).  A decrease in [Fe/H] of $-$0.5
is at most a decrease of 0.01 dex in A(Li), while the uncertainties in [Fe/H]
are typically much smaller at $\pm$0.06.  There is no change in A(Li) when
$\xi$ is changed by 0.2 km s$^{-1}$.  The uncertainty in $T_{\rm eff}$ is a
more important contributor where an increase of 100 K produces an increase in
A(Li) of 0.07-0.10 dex.  The uncertainties in $T_{\rm eff}$ are listed in
Table 5 of SB02 and these were used to find the error in A(Li) for each star.
One other contributor to the error is the measurement uncertainty in the Li
equivalent width.  As shown in SB02 for spectra with S/N $>$ 150 this error is
1 m\AA.  A conservative error of $\pm$2 m\AA\ gives an error of $\pm$0.025 in
A(Li).  The errors on A(Li) listed in Table 1 are the square root of the
quadrature sum of the uncertainties in A(Li) due to temperature and equivalent
width measurement uncertainties, and they are typically 0.06 dex.

For a subset of eight stars we have determined A(Li) from spectrum synthesis
with MOOG.  For seven of the stars the agreement is perfect.  For G 188-20 the
synthesis value of A(Li) is 0.02 dex lower than the equivalent width result;
this is certainly within the errors of the determination for this star of
0.05 dex.  The synthesis for the four stars in Figure 1 is shown in Figure 2.

According to the work of Carlsson et al.~(1994), there can be sizable
corrections to the Li abundances due to the effects of non-local thermodynamic
equilibrium (NLTE).  We have used the routines supplied by them to determine
the NLTE Li abundances given the LTE Li abundance, and the values of $T_{\rm
eff}$, log g, and [Fe/H].  The corrections range from $-$0.04 to +0.11 with
the negative corrections are for the hotter stars while the larger positive
corrections apply to the cooler stars.  For some of our stars one or more of
the four input values were out of the range of their calculations.  In some
cases we used log g = 4.5 when our value marginally exceeded that or [Fe/H] =
$-$3.0 for the three stars which were lower that that value.  In two cases
with low, but detected Li we extrapolated below the A(Li) = 0.6 limit in their
calculations.  The NLTE Li abundances are included in Table 1 where the values
for the stars with parameters beyond the calculation limits have a ``:'' after
the number to show that they are uncertain.  Most of the NLTE corrections
increase the LTE values by 0.01 to 0.11 dex (32 stars) while for six stars
they decrease the LTE values and for eight stars there is no change.  The
largest (positive) changes are for the cooler stars (5100 - 5200 K) with
intermediate metallicities ([Fe/H] = $-$1.34 to $-$1.90).  The small negative
corrections are for the hotter stars (6100 - 6300 K).  

\section{Results and Discussions}

\subsection{General Abundance Results}

Figure 3 shows the overall distribution of A(Li) in our star sample as a
function of temperature.  The hot stars in the sample are on the Li plateau
while the cooler star portion below 5700 K is heavily populated.  Solid
squares and triangles represent metal-poor ([Fe/H] $<$ $-$1.5) stars with Li
detections and upper limits, respectively.  Open squares represent the
metal-rich star Li detections.  This [Fe/H] dividing line isolates true
metal-poor halo stars, i.e.~those likely to possess a primordial Li abundance,
from stars whose surface abundance may be more affected by Galactic chemical
enrichment.  A further division was made in log g; the open circles have log g
$<$ 3.7, potentially subgiants.  This distinction is interesting when
discussing Li as the deepening convection zone of stars evolving off the main
sequence can penetrate to depths where Li is absent, resulting in the dilution
of the surface Li abundance.  Hence subgiant abundances may not represent an
unadulterated initial Li abundance.  Only five of our stars could be
considered to be subgiants, and three of those have temperatures below 5700 K.

The distribution of A(Li) with [Fe/H] is shown in Figure 4.  The stars plotted
as solid squares are the ones hotter than 5700 K with log g $>$3.7 and the
open squares are the cooler stars.  The circles refer to stars with log g $<$
3.7.  The two most metal-poor stars in the sample are {\it cool} subgiants: G
82-23 and G 238-30.  The star with the highest A(Li) is ``metal-rich'' at
[Fe/H] = $-$0.80 (G 121-12).  Stars that are Li-deficient are the {\it cooler}
ones, and occur at a range of metallicity.  The stars with upper limits
(triangles) occur at all metallicities, but only for the cooler stars.  We
have not discovered any additional ultra-Li deficient stars (at Li plateau
temperatures) in this sample.  Figure 5 puts our results into a larger context
of Li abundances in dwarfs.  The literature data are from Norris et
al.~(1997), the compilation of Ryan et al.~(1996), Ryan \& Deliyannis (1998),
Ryan et al.~(1999) and Mel\'endez \& Ram\'\i rez (2004).

There are 12 Li plateau stars with [Fe/H] $<-$1.5, T$_{\rm eff}$ $>$ 5700 K
and log g $>$ 3.7.  These stars have a mean A(Li) = 2.22 $\pm$0.12 similar to
traditional Li plateau abundances of $\sim$ 2.2 dex, e.g.~Bonifaco \& Molaro
(1997).  Other work has also focussed on the plateau stars.  The Ryan et
al.~(1999) study concentrated on the hottest and most metal-deficient halo
dwarfs, finding a very thin plateau and a low A(Li$_p$) of $\sim$2.0.  The
estimate of primordial Li by Ryan (2002) is 2.09 $\pm$0.16.  Zhang \& Zhao
(2003) find the Li plateau from 21 stars to be A(Li) = 2.33 and correct for a
metallicity dependence to derive A(Li$_p$) $\sim$2.08.  The recent
reevaluation of Li observations with revised temperatures (IRFM) by Mel\'endez
\& Ram\'\i rez (2004) yields A(Li$_p$) = 2.37, with an uncertainty in the
absolute abundance scale of 0.1.  Recently, Novicki (2005) has found both a
metallicity and a temperature dependence in the Li plateau from 116 stars;
after corrections for Li depletion and galactic chemical evolution, she finds
A(Li$_p$) = 2.44 $\pm$0.18.

We add two stars to our sample of 12 Li plateau stars which have log g = 3.54
and 3.53; these are the two stars with solid circles in Figures 3 and 4 and
have Li plateau abundances (G 88-32 and G165-39).  The mean for A(Li) for
these 14 stars is 2.215 $\pm$0.110.  As can be seen in Figure 4
there is a trend of A(Li) with metallicity.  The least squares fit of this
trend seen in Figure 6a is

A(Li) = 0.179 ($\pm$0.040) [Fe/H] + 2.649 ($\pm$0.099)

\noindent
Although the sample size is small, this does agree with other detections of a
metallicity dependence (e.g. Ryan et al.~1999, Spite et al.~2000, Novicki
2005).  Figure 6b shows the trend of Li with another Fe-peak element, [Cr/H];
the [Cr/H] values are also from SB02.  For [Cr/H] the slope of the
relationship is 0.177 $\pm$0.036, in excellent agreement with the slope
between [Fe/H] and A(Li).  A similar relationship is found for the other
Fe-peak element, [Ni/H]; that slope is a little shallower at 0.147 $\pm$0.040,
but the same within the errors.  Such trends could indicate that some chemical
evolution of Li has taken place even in these stars on extreme orbits.  On the
other hand, the lower Li in the lowest metallicity stars could result from
greater Li depletion in stars with lower quantities of the Fe-peak elements.
This could be expected if the stars with the lowest Fe are the oldest stars,
which would have had more time to deplete their original Li.  We do not find a
believable trend with temperature in this sample, i.e.~the slope per 100 K is
0.0121 $\pm$0.0215.

Our subgiant abundance data do not deviate from the data collected by
Pilachowski et al.~(1993) and Ryan \& Deliyannis (1995).  The three more
evolved subgiants (cooler temperatures) follow the typical pattern of
decreased Li with decreasing temperature, a trend attributed (in part) to main
sequence Li burning and subgiant Li dilution (Pilachowski et al.~1993).  The
two hotter stars with higher log g, G 88-32 and G 165-39, show the plateau Li
abundances (as mentioned above) of 2.17 and 2.20.

\subsection{Trends with Composition and Kinematics}

We have made a sub-sample of metal-poor stars ([Fe/H] $< -$1.5) with $T_{\rm
eff}$ $\gtrsim$ 5700 K to examine the potential trends with orbital
characteristics from SB02.  This subset contains only 14 stars and shows
values for A(Li) ranging from 1.97 to 2.45.  We have investigated trends with
orbital parameters for this subset.  The stars with large values (20-45 kpc)
of $R_{\rm apo}$ (the distance at apogalacticon) have the same mean value of
A(Li) as those of with $R_{\rm apo}$ of 8-12 kpc.  The orbital energy, $V_{\rm
RF}$, is similar for the five stars with the highest A(Li) and those with
lower A(Li).  This subset has a mean A(Li) = 2.22 $\pm$0.11.  Eight of those
14 stars are from the outer halo and show essentially the same A(Li): 2.25
$\pm$0.10.  Four stars are from the high halo and have A(Li) = 2.18 $\pm$0.07,
similar to the total sample of 14 stars.  The small difference between outer
halo stars and high halo stars is not significant, nor are either different
from the halo samples of other research on halo stars on ``normal'' orbits.
That we have found no important differences is not surprising given the
primordial origin of Li.

Table 2 lists the stars in the plateau subset of 14 stars along with their
parameters and the abundances and errors of several elements.  This group is
the true halo plateau sample with T$_{\rm eff}$ $>$ 5700 K, log g $>$ 3.5 and
[Fe/H] $<$ $-$1.5.  The values of [Mg/H], [Ca/H] and [Ti/H] were used as
indicators of alpha-element enrichment.  (Only 2 of the 14 stars have
measurements of the other alpha-element, Si.)  The abundance of Li increases
in these stars with all three of the $\alpha$-element ratios: [Mg/H], [Ca/H]
and [Ti/H].  Figure 7a and 7b show the increase with Mg and Ca and Figure 8a
shows the increase with Ti.  The slopes of these three relationships are
remarkably similar at 0.216 $\pm$0.047 for Mg, 0.202 $\pm$0.042 for Ca, and
0.201 $\pm$0.043 for Ti.  These slopes are higher than those of the three
Fe-peak elements as would be expected because of the super-solar values of
[$\alpha$/Fe] in old, metal-poor stars and the increasing rise to higher
[$\alpha$/Fe] with decreasing [Fe/H] as found in SB02.

In addition we looked for trends of A(Li) with the neutron-capture elements.
Although Y was only measurable in four of the 14 stars, we have abundances of
Ba in all of them.  This trend is shown in Figure 8b.  In the case of [Ba/H]
the slope is small: 0.130 $\pm$0.032.  The material that formed these halo
stars was apparently not exposed to as much enrichment of s-process products
as to the $\alpha$-process products.  The increase in A(Li) is only a factor
of about 2 over the span of 2.6 dex in [Ba/H].

It is interesting to observe how much chemical enrichment has occurred in
these halo stars and how the apparent Li enrichment has kept pace.  Quite
different mechanisms are involved as Li is produced by spallation
and the others by stellar nucleosynthesis.  However, we see no particular
evidence that there is any Li over-enrichment in these stars that could be
attributed to the $\nu$-process.  Furthermore, the possibility remains that we
are not seeing evidence of Li enrichment by spallation, but rather evidence of
Li depletion which is larger in the lower metallicity (older?) halo stars.

\subsection{Cool Metal-Poor Stars}

The data on the non-plateau stars exhibit some interesting new trends.  This
dataset is unique in that it contains a large number of metal-poor stars with
effective temperatures cooler ($T_{\rm eff}$ $<$ 5700 K) than the Li plateau.
Figure 9 shows examples of the spectrum synthesis of Li in four stars with
similar metallicities in which the Li line strength and A(Li) decline with
decreasing temperature.  The best synthesis fits are shown along with
syntheses which are a factor of two more and two less Li.

It is clear from Figure 3 that the cool, metal-poor stars (solid squares)
possess, on average, more Li than stars with similar temperatures and larger
metallicities (open squares).  The metal-richer stars appear to fall along a
``lower envelope'' while the metal-poorest stars appear to define an
``upper-envelope'' to the Li depletion curve over the entire range of
temperatures cooler than the plateau.  This result agrees with similar
findings based on much less data (Deliyannis, Pinsonneault, \& Duncan 1993;
Ryan \& Deliyannis 1995; and Ryan \& Deliyannis 1998).

In order to examine the trends of Li depletion with metallicity in the cool
stars in more detail, we divide the stars into different metallicity
sub-groups.  We have excluded stars with log g $<$ 3.7, stars hotter than
5800 K, and those with only upper limits on A(Li).  The resulting sample of 22
stars has been subdivided into three metallicity groups: [Fe/H] = $-$0.70 to
$-$1.03; $-$1.24 to $-$1.47; and $-$1.58 to $-$1.90.  (Of the remaining six
stars, three fall between these groups.  One star, G 97-40, was not plotted
because at [Fe/H] = $-$1.52 it is intermediate between our metallicity
groupings, but at [5427,1.20] it is in reasonable accord with the [Fe/H] =
$-$1.24 to $-$1.47 group.  Three have lower metallicity and could fit with our
lowest metallicity group, but that would extend the range in that group to
[Fe/H] to $-$2.14.)  The stars in the three groups are listed in Table 3.  All
have log g $>$ 4.14.

The trends with metallicity become even more apparent in Figures 10 and 11.
The decline in A(Li) with temperature is similar in the three groups, but it
is offset toward cooler temperatures for the lower metallicity groups.  The
lines shown in Figure 10 are the least squares fits through the points with
two such fits for the two lower metallicity groups.  (The outlier, G189-050,
at [5254,1.46] was excluded from the fit.  The most likely source of error is
the value of [Fe/H]; if this were lower by 0.1 dex it would be included in the
lower metallicity grouping, where it lies in Figure 10.)  The change in the
slopes occurs near A(Li) $\sim$1.1.  The slopes for the upper parts are 2.66
($\pm$0.02) $\times$ 10$^{-5}$, 2.21 ($\pm$0.08) $\times$ 10$^{-5}$, and 2.06
($\pm$0.84) $\times$ 10$^{-5}$ for $<$[Fe/H]$>$ = $-$0.86, $-$1.41, and
$-$1.74, respectively.  These are very similar slopes, but the same Li
depletion occurs at {\it cooler} temperatures for the {\t lower metallicity}
stars.  At a given temperature, e.g. 5400 K, there is more than 4 times as
much Li depletion in the more metal-rich stars.  The slopes steepen for the
cooler, more metal-poor stars.

Figure 11 shows the same type of plot, but it is for the NLTE A(Li) values.
(The lines shown in Figure 11 are not ``fits'' to the data, but rather eyeball
connect-the-dots lines.)  In that diagram the Li abundances are systematically
higher resulting in an apparent shift toward the right for the
connect-the-dots lines.

Although the error bars are shown for both T$_{\rm eff}$ and A(Li) in the
figures, the question arises about systematic differences in temperature
scales.  Of the 16 stars in Table 3 there are 5 in common with the Alonso et
al.~(1996) temperature determinations.  Within the errors the temperatures for
four of the stars agree with ours, but the errors in the Alonso determinations
in these cool stars are large: $\pm$108 K to $\pm$206 K with a mean of 152 K.
(Our errors for these five temperatures are 40 - 89 K with a mean of 63 K.)
Five of the stars in Table 3 have large reddening corrections which makes
deriving photometric temperatures more problematic, but we can compare the
temperatures of the other 11 stars with those of Carney et al.~(1994).  The
temperature differences (SB02 $-$ Cetal) range from -69 to +205 K.  While our
temperatures are generally hotter than those of Carney et al.~for these cool
stars, the effect of using cooler temperatures would be to decrease the Li
abundance.  This generally moves the data points in Figures 10 and 11 along
the lines drawn, and may flatten the slopes somewhat.  For a change of $-$100
K the value of A(Li) is decreased by 0.10 dex.

{\it The decline in Li abundance sets in at higher temperatures for the
metal-rich stars than for the metal-poor stars in both Figures 10 and 11.}
This may result from a simplified view of the extent and influence of
convection in metal-poor stars.  The surface convection zone deepens with
decreasing temperature which causes the depletion of Li in cool stars as Li
gets mixed to deeper, hotter layers by convection currents and destroyed there
by nuclear reactions.  In low-metal stars there is less internal opacity which
results in shallower convection zones in those stars with lower metal content.
As we have shown in Figures 10 and 11 there is less Li destruction at a given
temperature in the metal-poor stars and a given depletion of Li occurs at
cooler temperatures in metal-poor stars.

However, the results in Figures 3, 10, and 11 are in stark contrast with the
actual predictions of ``standard'' stellar models of light element depletion.
Such models ignore diffusion, rotation, mass loss, and magnetic fields and
predict that metal-poor stars should experience more depletion than metal-rich
stars at the same $T_{\rm eff}$ (Deliyannis, Demarque, \& Kawaler 1990;
Deliyannis \& Demarque 1991).  Figure 12 shows, on the same scale, the
predictions from the models for 16.5 Gyr of Li depletion tracks for [Fe/H] =
$-$1.5 and $-$2.3 from the Yale ``standard'' models (from Ryan \& Deliyannis
1998).  Not only do the models predict more Li depletion rather than less for
the metal-poorest stars, they also predict a steeper decline in Li abundance
than is observed.  These standard model predictions fail to reproduce the
abundance trends.

The depth of the convection zone in the models is nearly independent of
metallicity for a given temperature.  So the metallicity dependence that we
have found (Figure 10) points out that the mixing effects are poorly
understood for low mass halo dwarfs.  There may be one aspect where the models
marginally agree.  Halo star models which track instabilities due to
superficial spin-down and angular momentum loss (Deliyannis, Pinsonneault, \&
Duncan 1993; Pinsonneault, Deliyannis, \& Demarque 1992, Pinsonneault et
al.~1999) predict a larger Li {\it dispersion} should be present in stars with
increasing metallicity (at a given $T_{\rm eff}$).  It is not clear from
Figures 3 and 5 if there really is a larger Li dispersion in the higher
metallicity stars (open symbols) compared to the lower metallicity ones (solid
symbols).  And such a difference in dispersion could also be due to Galactic
Li enrichment, perhaps seen in Figures 6-8 of increasing Li with increasing
Fe, Cr, Mg, Ca, Ti, and Ba.

Those models which include rotationally-induced mixing are surely more
realistic `` standard models.''  There are other lines of
evidence suggest that rotationally-induced mixing is a plausible cause of the
observed light element abundance patterns in metal-poor stars.  In particular,
the correlation between A(Li) and A(Be) in field and cluster disk stars is
best explained by rotationally-induced mixing (e.g.~Boesgaard et al.~2004).
Also, Li is enhanced at a given $T_{\rm eff}$ in short-period tidally-locked
binaries which are thought to experience little rotationally-induced mixing
(Ryan \& Deliyannis 1995).  And trends with $T_{\rm eff}$ and [Fe/H] (Thorburn
1994, Norris, Ryan, \& Stringfellow 1994, Ryan et al.~1996, Ryan et al.~1999)
in the Li plateau may be caused by internal mixing process affects the
superficial light element concentrations in metal-poor stars.

\section{Summary and Conclusions}

We have obtained spectra of 55 metal-poor stars, primarily with the Keck I
telescope with HIRES at high S/N per pixel (70 - 700) and high spectral
resolution ($\sim$48,000) and have determine Li abundances in 42 stars and
upper limits on Li in the other 13.  The Li abundances were found with the
assumption of local thermodynamic equilibrium and then corrected for NLTE
effects.  

For our 14 stars in the Li plateau region ($T_{\rm eff}$ $\gtrsim$5700 K) with
[Fe/H] $<$ $-$1.5 and log g $>$ 3.5 we find A(Li) = 2.215 $\pm$0.110.  For
this subset we find trends in A(Li) with the Fe-peak elements, Cr, Fe, and Ni,
with similar slopes near 0.15 to 0.18.  In addition, there are clear
correlations between A(Li) and the $\alpha$-elements, [[Mg/H]. [Ca/H], and
[Ti/H].  For these elements the slopes are somewhat greater than those for the
Fe-peak elements at 0.20 to 0.21.  This is to be expected because of the
super-solar ratios of alpha-elements in metal-poor stars which rise to higher
values of [$\alpha$/Fe] with decreasing [Fe/H], as shown in SB02.  The
neutron-capture element ratio, [Ba/H] is also correlated with A(Li) with a
shallower slope of 0.13 possibly indicating that these low metallicity stars
have not been exposed to much enrichment by the s-process.  The observed
correlations could result from Galactic chemical enrichment of these elements
with time, or be the result of slow Li depletion with preferentially greater
depletion occurring in successively more metal-poor stars.  We examined this
dataset for Li abundance patterns with orbital characteristics, e.g.~total
orbital energy, distance at apogalacticon, stars from the high halo, or on
retrograde orbits, and found no significant differences.

The majority of our stars are cool, metal-poor dwarf stars and these stars
exhibit interesting trends of A(Li) with $T_{\rm eff}$.  As with metal-rich
disk dwarfs, A(Li) declines with temperature as the surface convection zone
deepens and Li is depleted as it is mixed to high enough temperatures to
destroy it.  Our stars with [Fe/H] $<-$1.4 form an upper envelope of this
depletion profile.  We have looked at this in more detail with three
metallicity subsets with mean [Fe/H] values of $-$0.9, $-$1.4, $-$1.7.  The
depletion sets in at cooler temperatures for successively metal-poorer stars.
It is expected that the convection zones are not as extensive for the lower
metallicity stars due to their lower internal opacity.  Yet the observations
are in contrast to the predictions for Li depletion from standard ``Yale''
models.  Nor do the models with rotational spin-down agree with the sense or
the size of the depletion with metallicity.  The errors in A(Li) are typically
$\pm$0.06 and 70 K for $T_{\rm eff}$ so the observations seem secure.

\acknowledgments 
This work has been supported by NSF grant AST 00-97945 to AMB.

\clearpage

\begin{figure}
\plotone{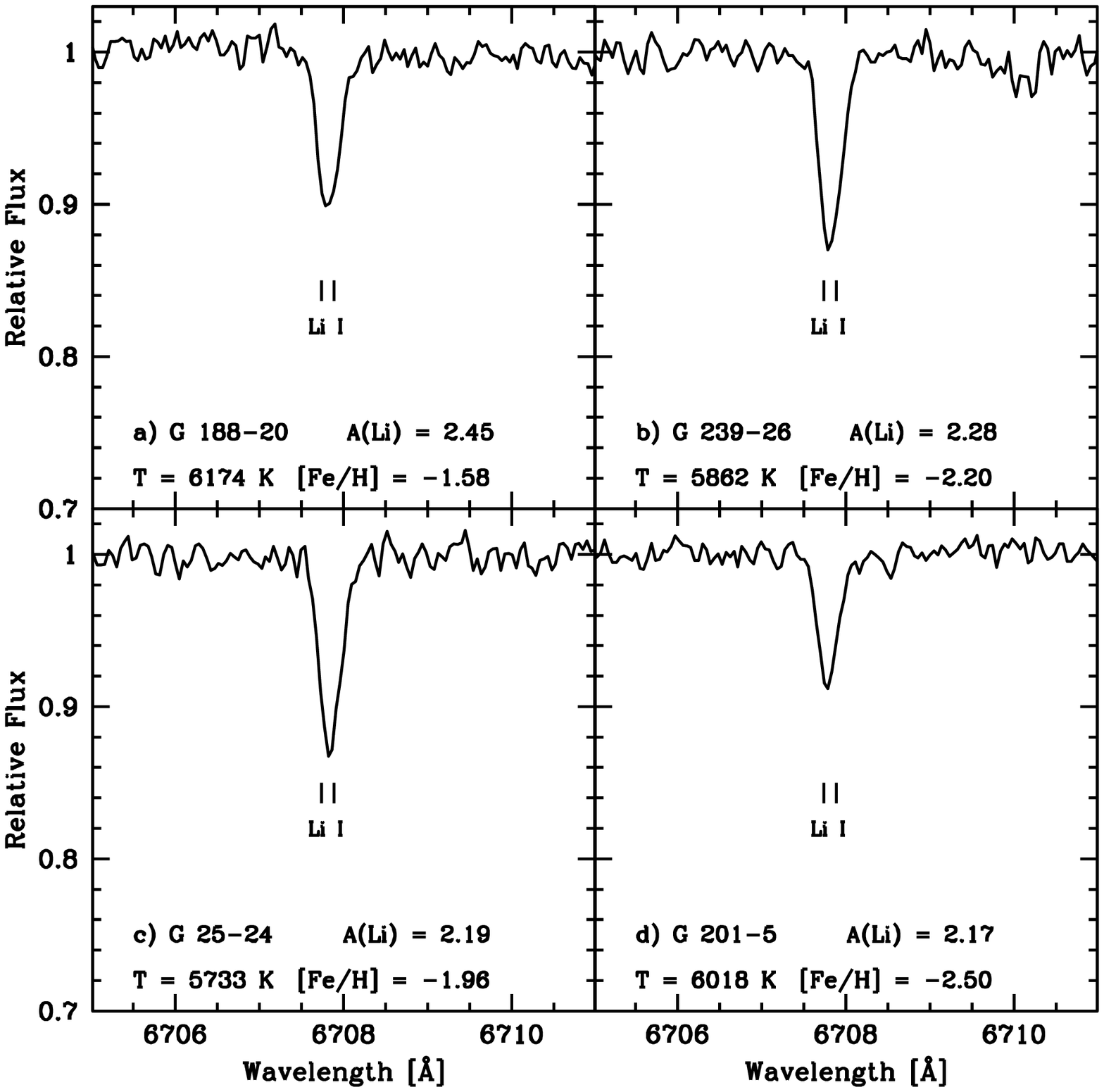}
\caption{Examples of the spectra in the Li region for some of the Li-plateau
stars in our sample.  Three are from the outer halo (a, b, d) and two from
high in the halo (b,c).}
\end{figure}

\begin{figure}
\plotone{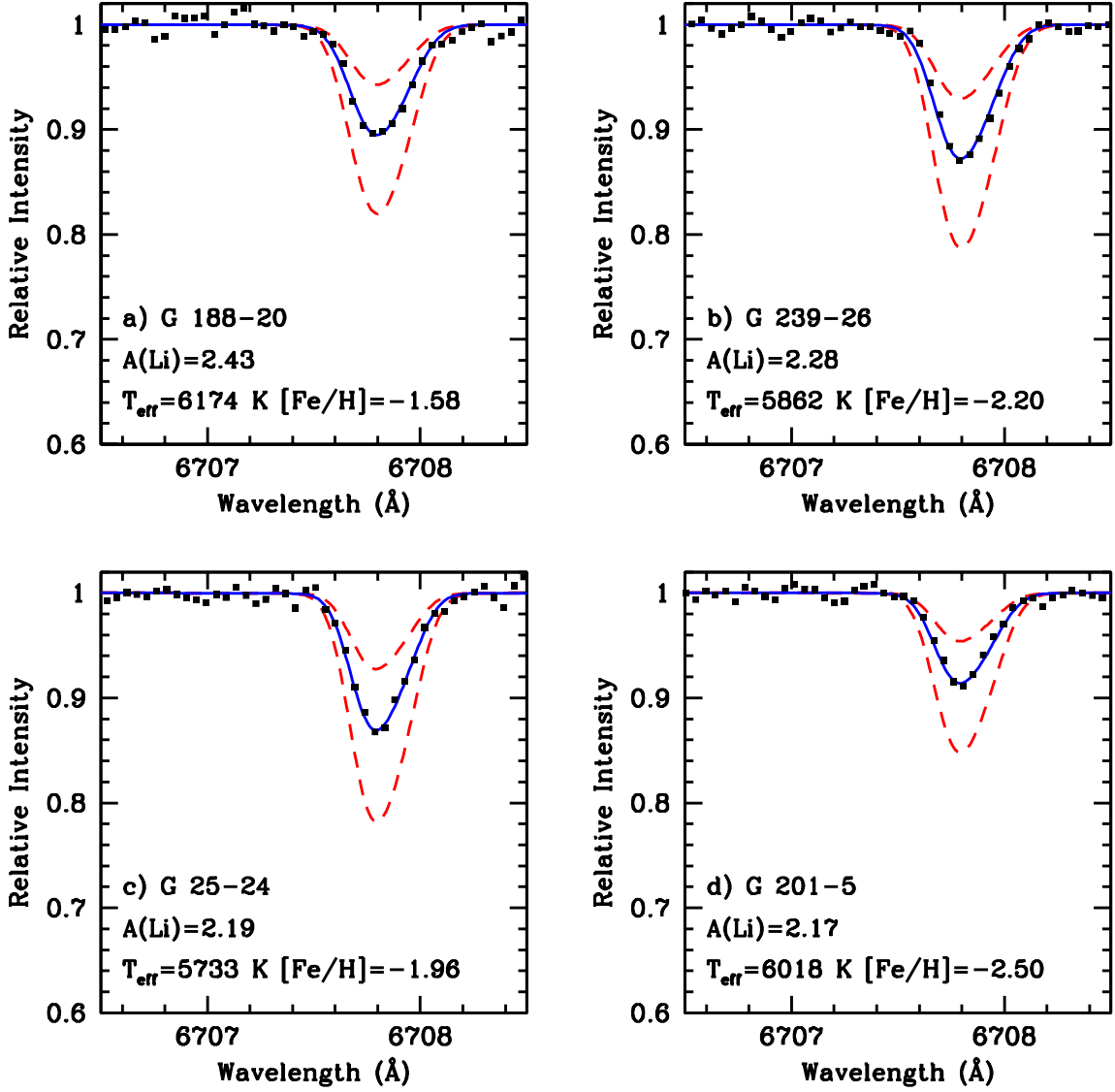}
\caption{Spectrum synthesis calculations for the spectra of the stars shown in
Figure 1.  The observations are shown by the small squares, the best fit is
the solid line, while the dashed lines correspond to a factor of two more Li
and a factor of two less Li.}
\end{figure}

\begin{figure}
\plotone{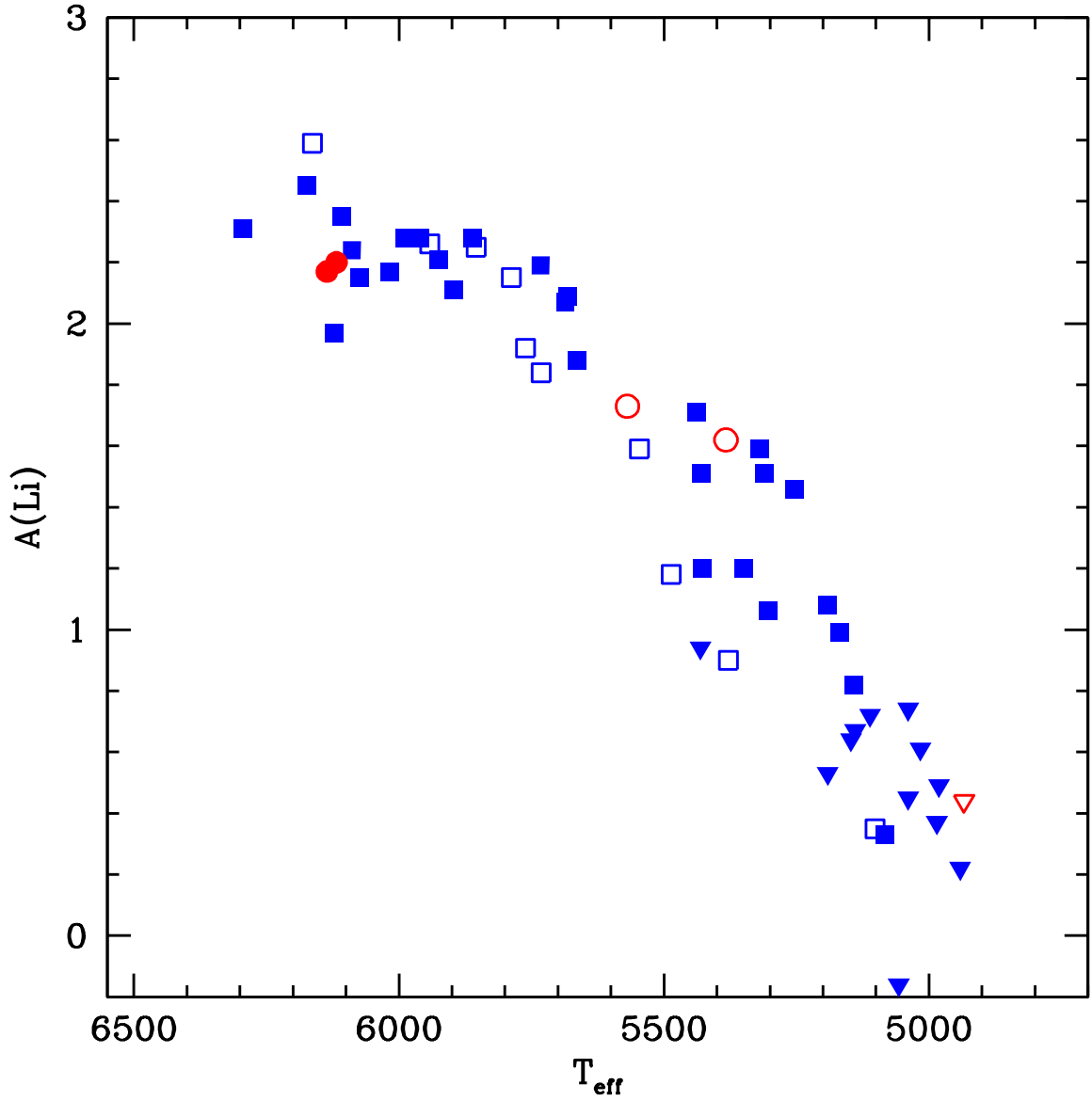}
\caption{The distribution of Li abundances with temperature.  Solid squares
and triangles are dwarf stars with low metallicity, [Fe/H] $<$ $-$1.4, while
the open squares are stars with higher metallicity.  The solid circles, open
circles and a triangle are stars with log g $<$ 3.7.}
\end{figure}

\begin{figure}
\plotone{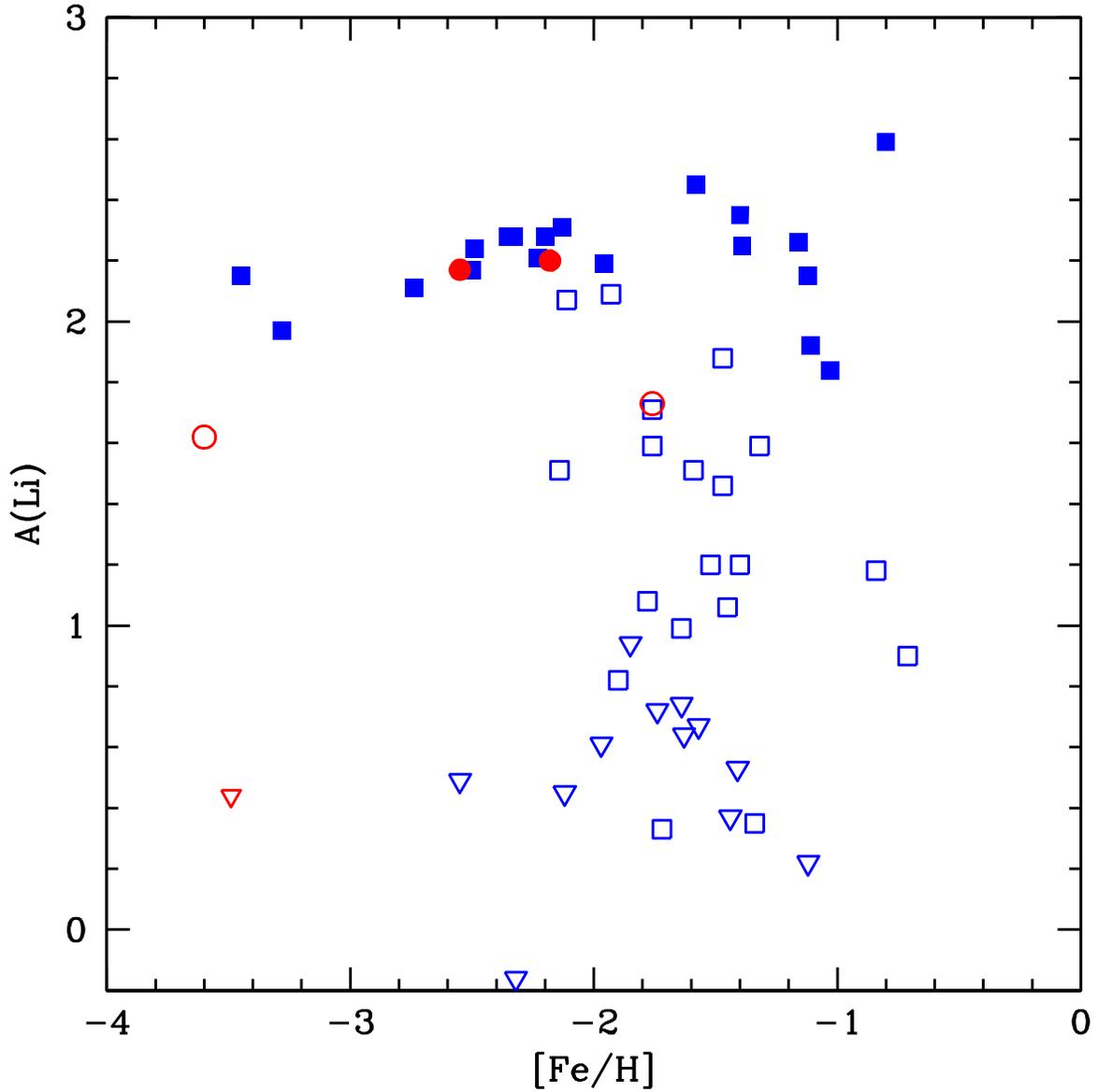}
\caption{The distribution of Li abundances with [Fe/H].  Solid squares are
dwarf stars with T$_{\rm eff}$ $>$ 5700 K while the open squares have lower
temperatures.  The open triangles are the cool stars with upper limit Li
abundances.  The circles and one triangle are stars with log g $<$ 3.7 where
the filled circles are the hotter stars and the open circles the cooler
stars.}
\end{figure}

\begin{figure}
\plotone{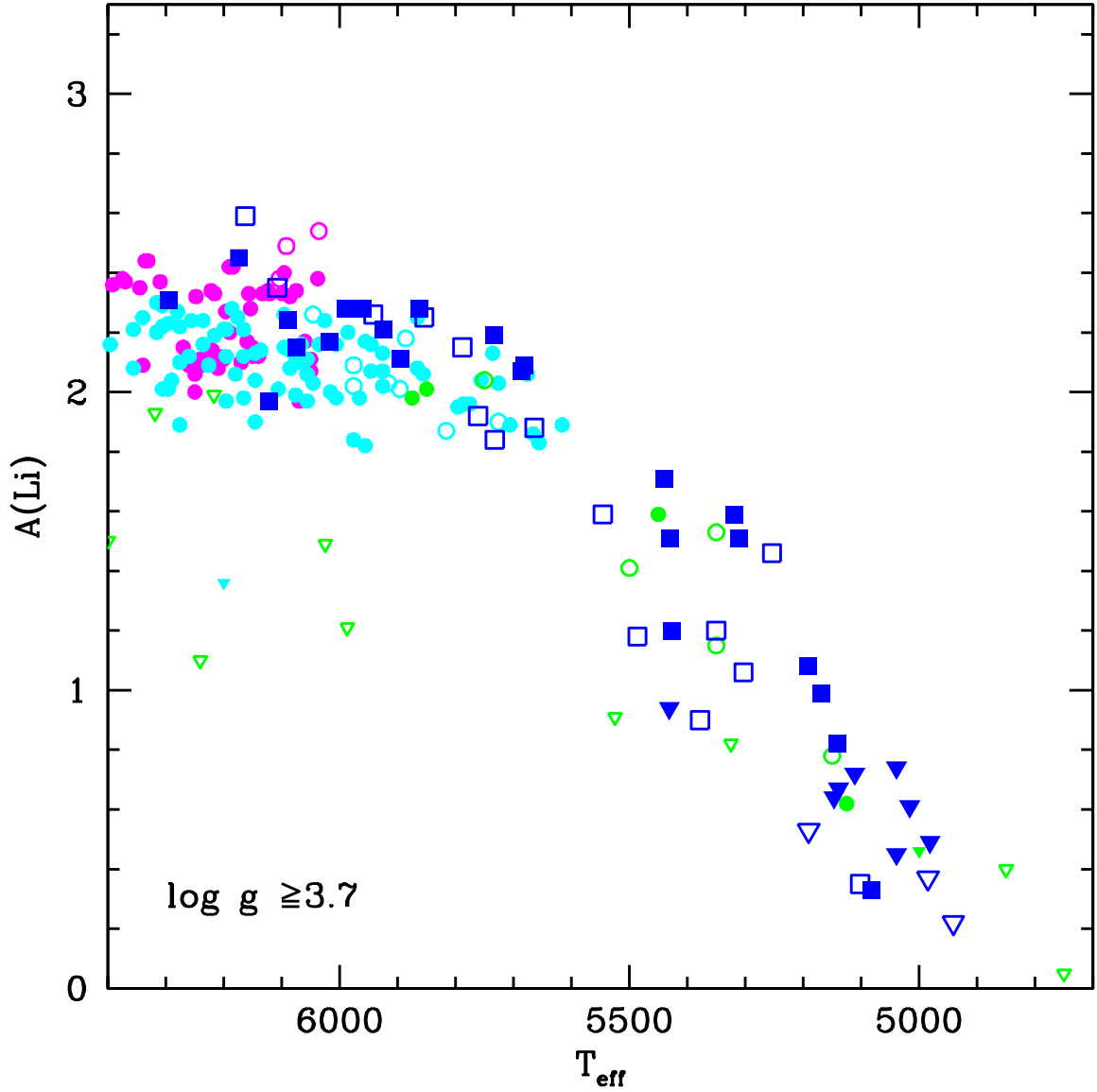}
\caption{The same as Figure 3, except Li abundances from the literature are
superimposed upon the plot.  The small circles and triangles represent dwarf
star data from Norris et al.~(1997), Ryan et al.~(1996), Ryan \& Deliyannis
(1998), Ryan et al.~(1999) and Mel\'endez and Ram\'\i rez (2004).  Open
symbols are for higher metallicity stars: [Fe/H] $>$ $-$1.5}
\end{figure}

\begin{figure}
\plottwo{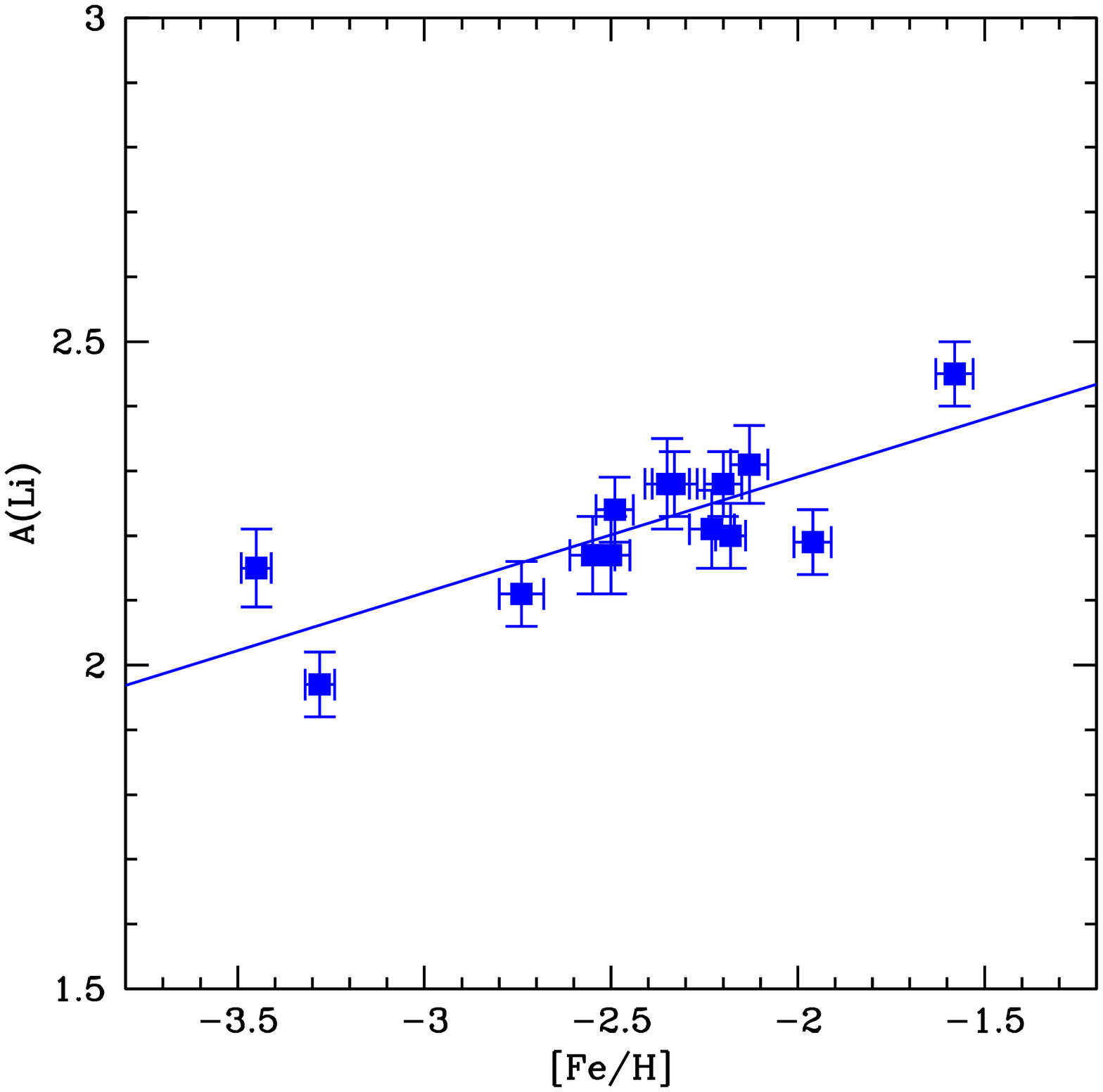}{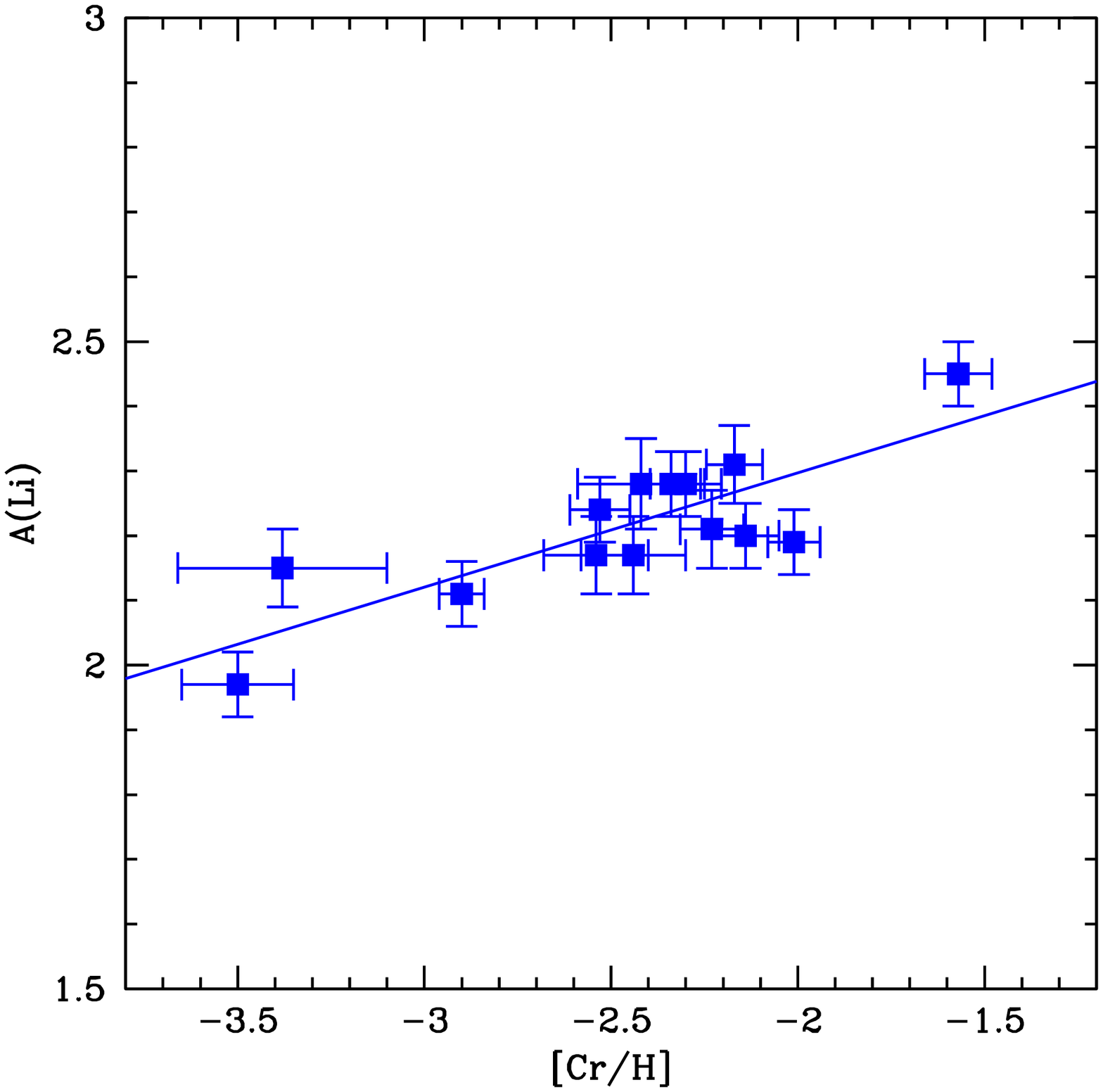}
\caption{The trend of Li abundance with the Fe-peak elements, [Fe/H] (left)
and [Cr/H] (right) for the 14 Li plateau stars.  The slopes of these
relationships are 0.179 $\pm$0.040 ([Fe/H]) and 0.177 $\pm$0.036 ([Cr/H]).
Even though these halo stars are on extreme Galactic orbits, this may
represent evidence for chemical evolution effects.  Alternatively, it could
indicate that there is greater Li depletion in the more metal-poor stars.}
\end{figure}

\begin{figure}
\plottwo{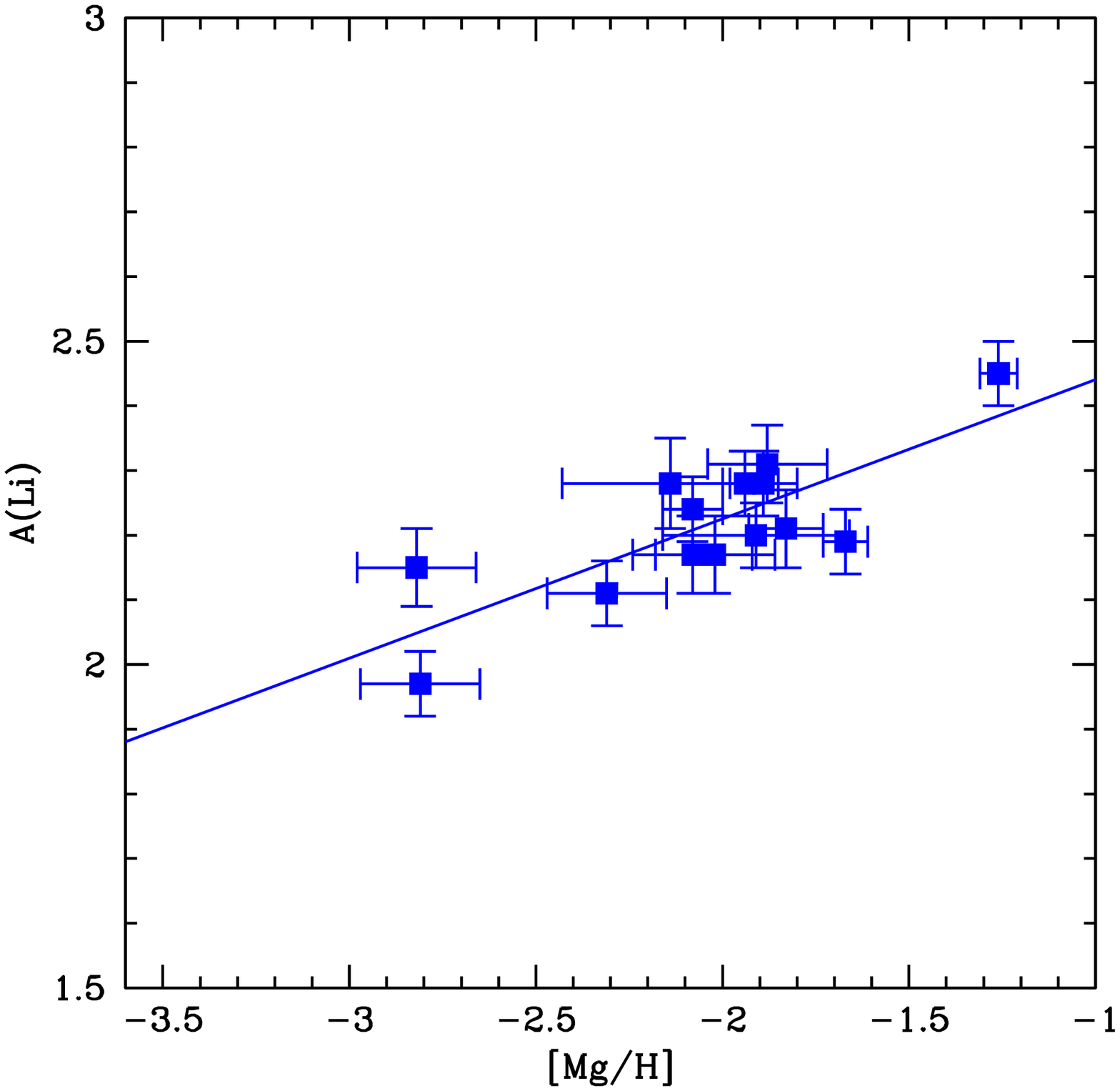}{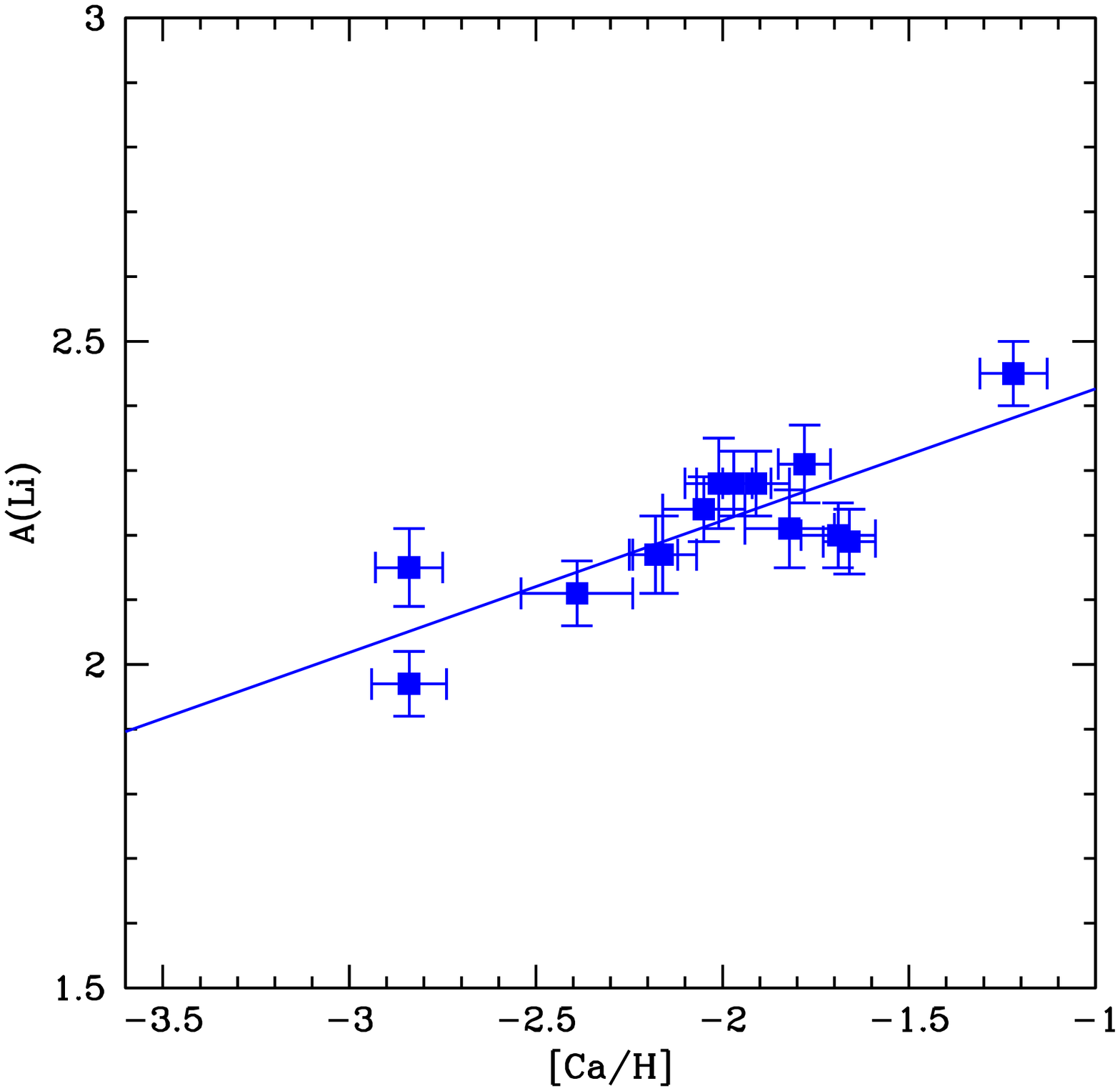}
\caption{The trend of Li abundance with the $\alpha$-elements, [Mg/H] (left)
and [Ca/H] (right) for the 14 Li plateau stars.  The slopes of these
relationships are 0.216 $\pm$0.047 ([Mg/H]) and 0.204 $\pm$0.042 ([Ca/H]).
These slopes are larger than the Fe-peak elements, indicating the greater
relative production of $\alpha$-elements in the early halo.}
\end{figure}

\begin{figure}
\plottwo{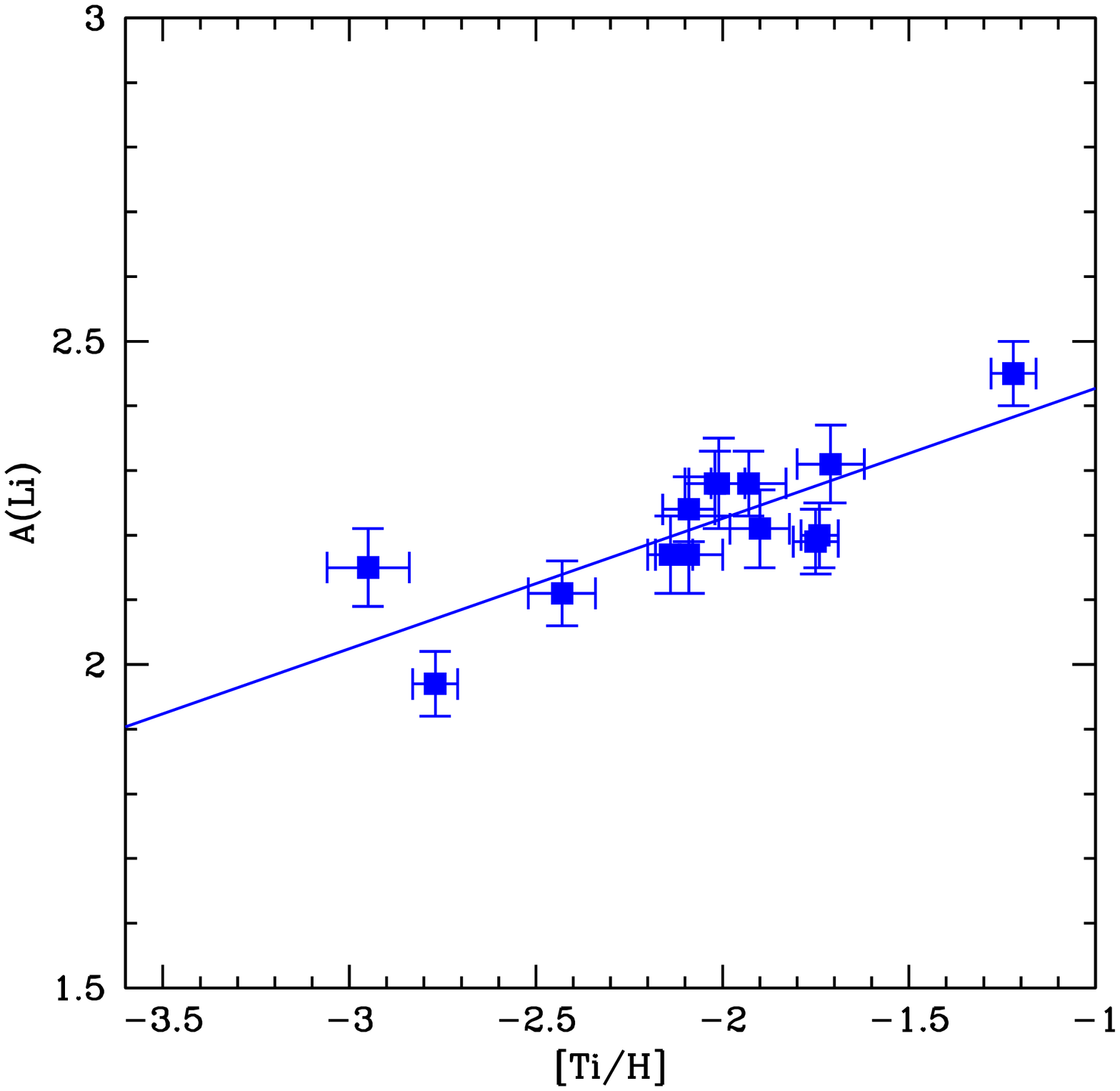}{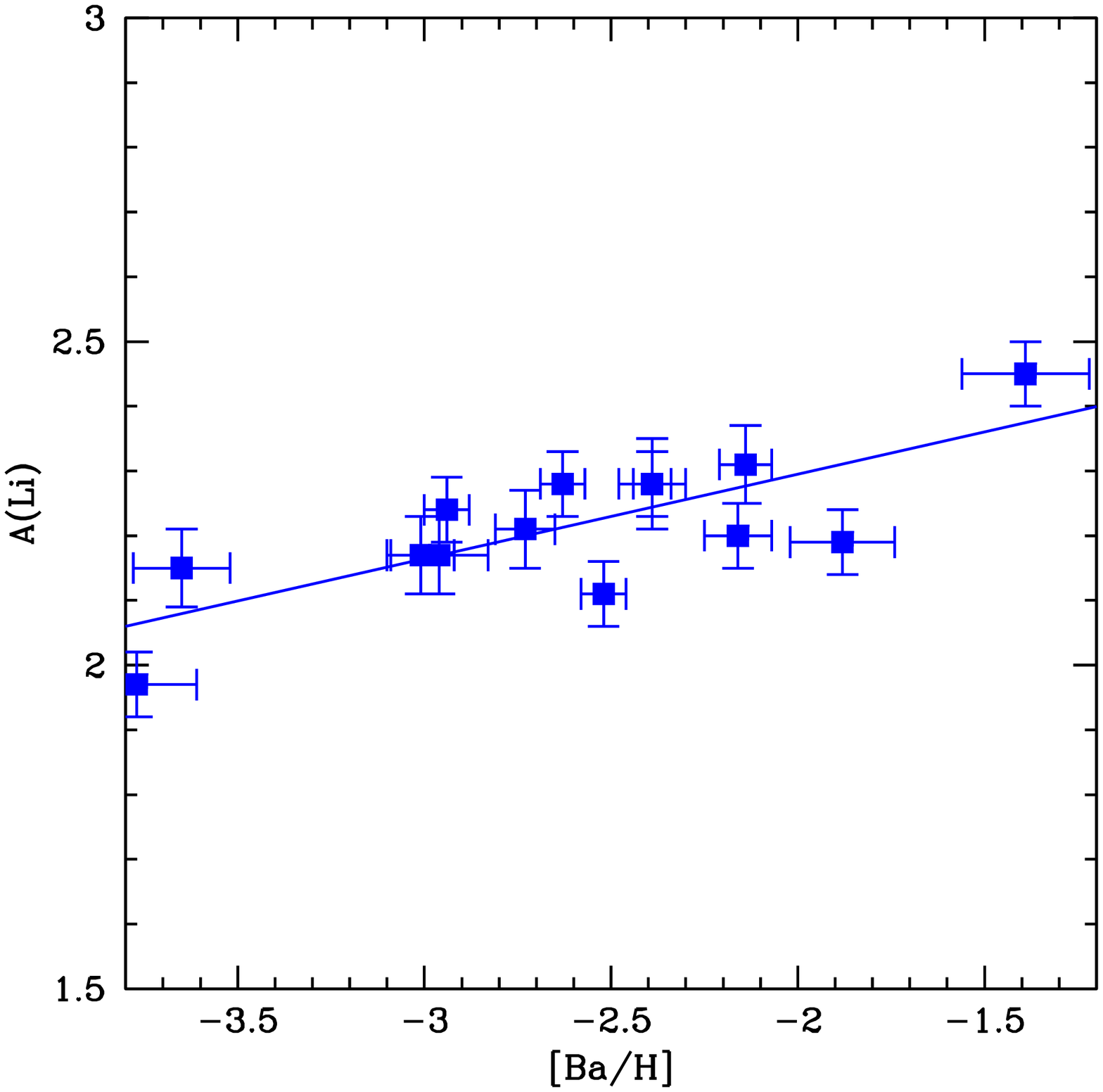}
\caption{The trend of Li abundance with [Ti/H] (left) and [Ba/H] (right) for
the 14 Li plateau stars.  The slope of our third $\alpha$-element, [Ti/H], is
0.201 $\pm$0.043, in remarkable agreement with the other two
$\alpha$-elements.  For the s-process element, [Ba/H], the slope is shallower
at 0.130 $\pm$0.032 indicating less neutron-capture activity.}
\end{figure}

\begin{figure}
\plotone{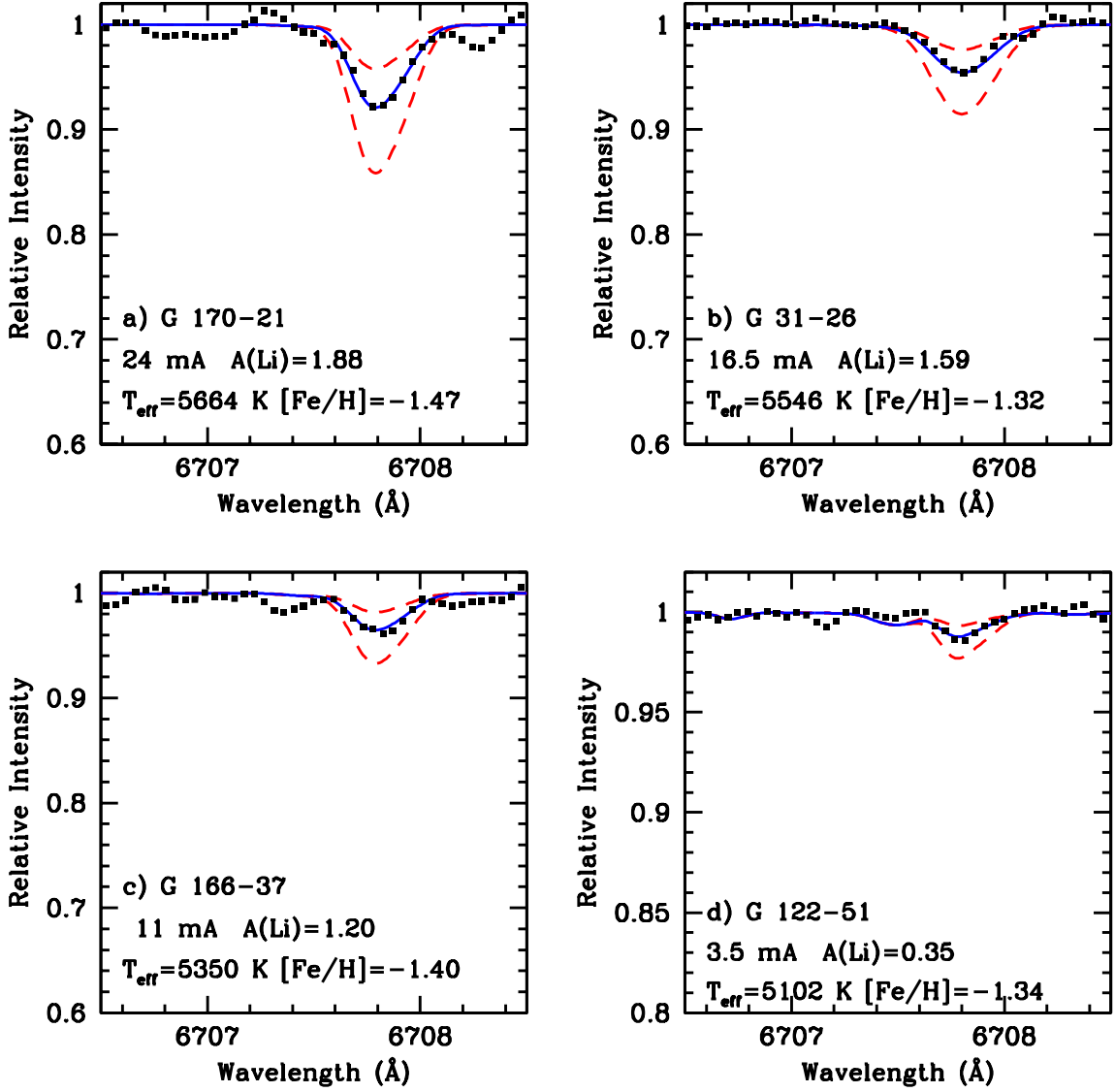}
\caption{Spectrum syntheses of four of the cool stars of similar [Fe/H] but
decreasing temperature and Li abundances.  The observations are shown by the
small squares, the best fit is the solid line, while the dashed lines
correspond to a factor of two more Li and a factor of two less Li.  These four
spectra correspond to the stars in the middle metallicity region ($<[Fe/H]>$ =
$-$1.41) in Figure 10.  The Li equivalent width, and A(Li), decrease as the
temperature decreases.  Notice there is a change in the vertical scale in
panel d.}
\end{figure}

\begin{figure}
\plotone{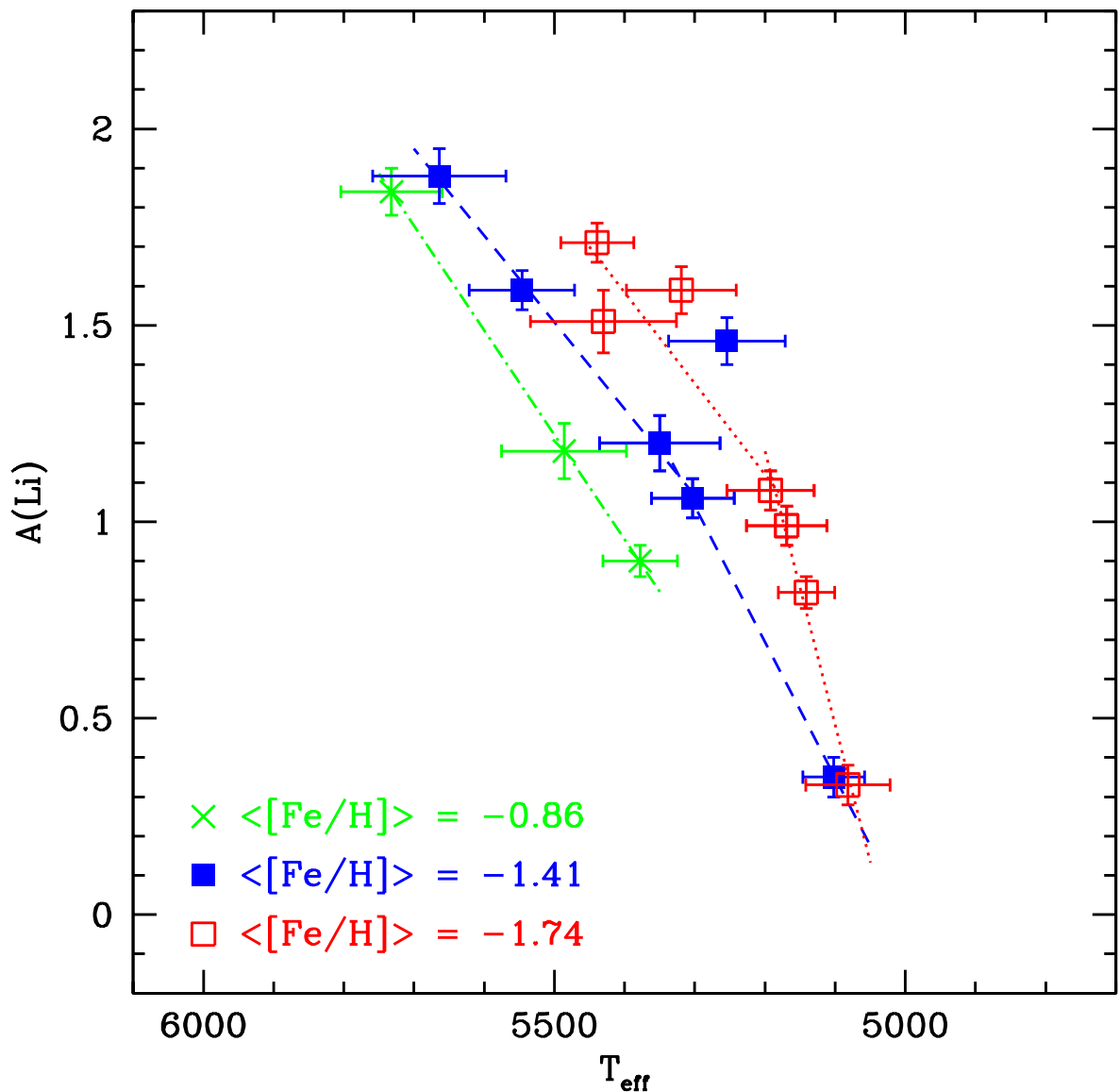}
\caption{Lithium abundances plotted against temperature in different
metallicity subgroups.  This shows the LTE value of A(Li).  The lines shown
are least-squares fits to the data with the two lower metallicity groups being
fit by two straight lines.  (The outlier at [5254,1.46] has been omitted from
the fit.)  The three metallicity groups show declining relationships with
temperature that are offset from one another such that the decline for lower
metallicity stars sets in at cooler temperatures.}
\end{figure}

\begin{figure}
\plotone{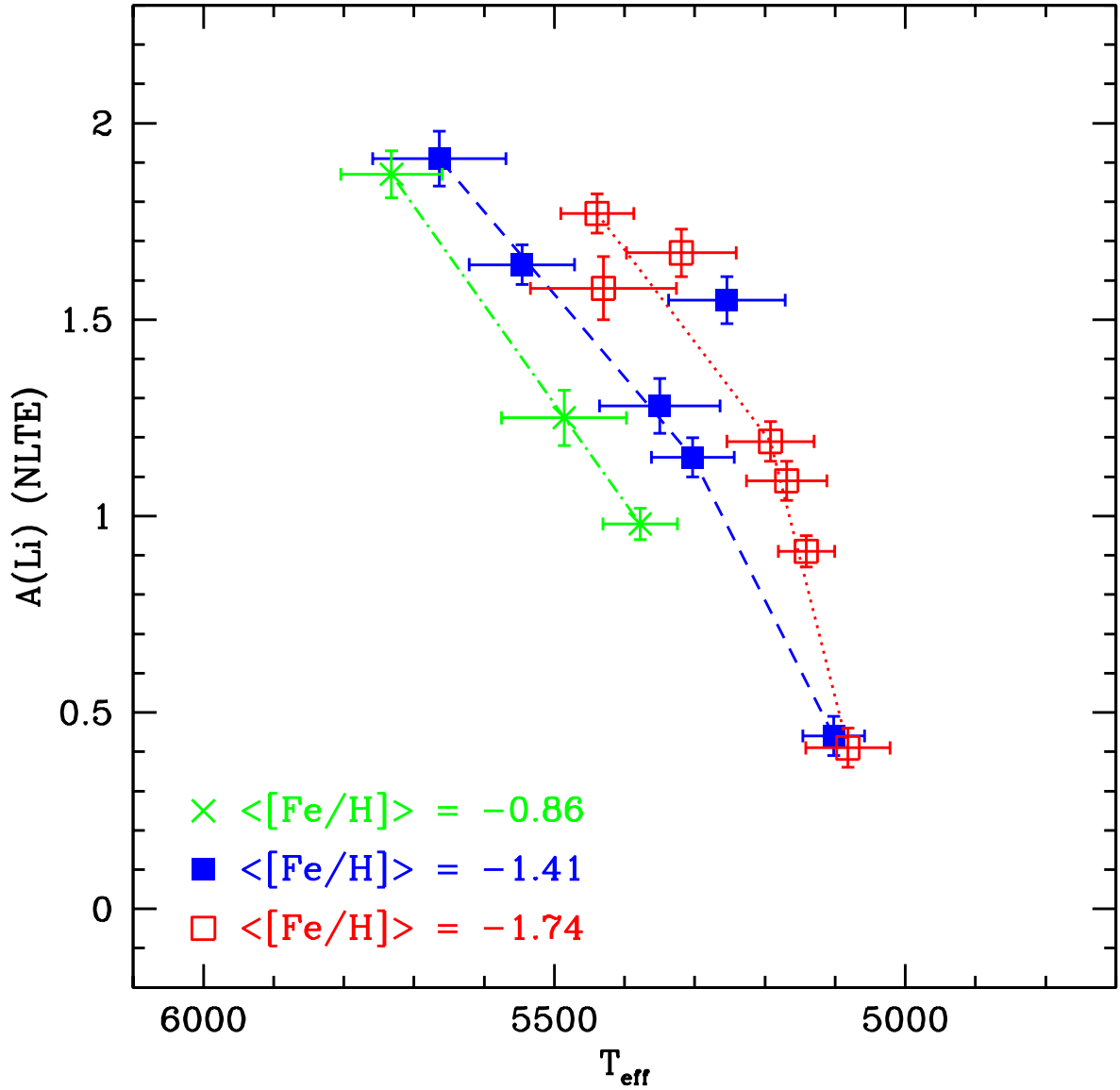}
\caption{Like Figure 10 but for the NLTE values of A(Li) and the lines shown
are not fits to the data, but rather ``connect-the-dots'' lines; these are
very similar to the fits in Figure 10.  Most of the NLTE A(Li) values increase
which has the effect of shifting the connecting lines to the right.}
\end{figure}

\begin{figure}
\plotone{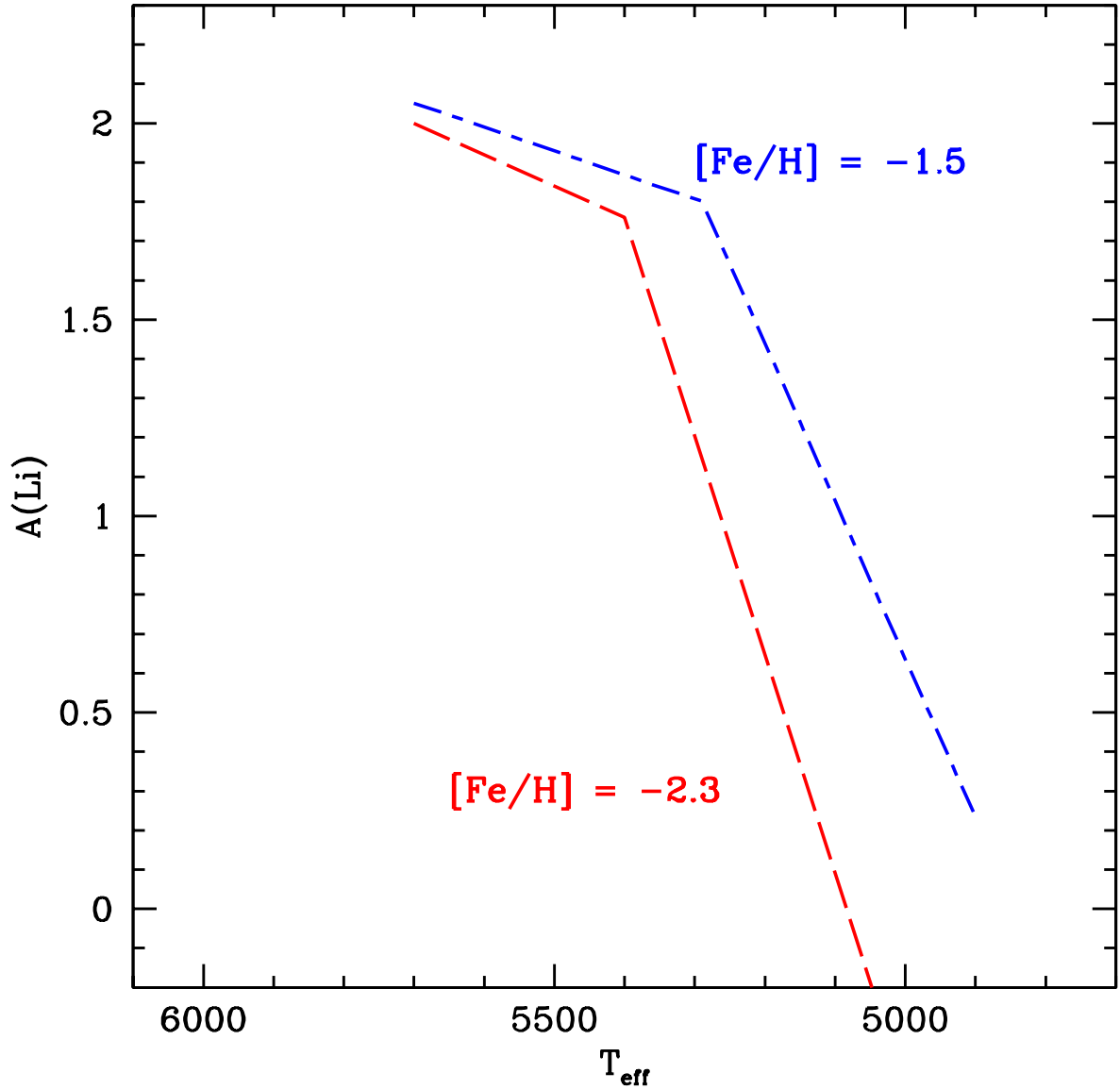}
\caption{Standard ``Yale'' model predictions for the Li-depletion tracks at
two metallicities for a 16.5 Gyr star, plotted on the same scale as Figures 10
and 11.  Note that in the model predictions the decrease sets in at higher
temperatures for the lower metallicity case, exactly the opposite of the
observations in Figures 10 and 11.}
\end{figure}

\clearpage
%\documentclass[12pt,preprint]{aastex}
%\documentstyle[apjpt4]{article}

%\begin{document}

\begin{deluxetable} {crcccrrrc}
%\footnotesize
\tablecolumns{9}
\tablewidth{0pc}
\tablenum{1}
\tablecaption{\  Stellar Parameters and Lithium Abundances }
\tablehead{
\multicolumn{1}{c}{Star}    &
\multicolumn{1}{c}{S/N}    &
\multicolumn{1}{c}{$T_{\rm eff}$}    &
\multicolumn{1}{c}{log g}    &
\multicolumn{1}{c}{[Fe/H]}    &
\multicolumn{1}{r}{Li EQW}    &
\multicolumn{1}{r}{A(Li)}   &
\multicolumn{1}{r}{A(Li)$_{NLTE}$}   &
\multicolumn{1}{r}{$\sigma$A(Li)}   
}
\startdata
%STAR    & S/N & Teff & grav & [Fe/H]  & EQW   & A(Li) & A(LI)NLTE & sig
G031-026 & 145 & 5546 & 4.50 & $-$1.32 & 16.5$\pm$1.0  & 1.59 & 1.64 & 0.05 \\ 
G171-050 & 185 & 5896 & 4.03 & $-$2.74 & 28.5$\pm$0.8  & 2.11 & 2.13 & 0.05 \\
G158-100 & 135 & 4981 & 4.16 & $-$2.55 & $<$4.4$\pm$1.1 & $<$0.49 &\nodata &\nodata \\
G033-031 & 130 & 4941 & 4.45 & $-$1.12 & $<$4.5$\pm$1.1 & $<$0.22 &\nodata &\nodata \\ 
G005-019 & 110 & 5942 & 4.24 & $-$1.16 & 33.0$\pm$1.3  & 2.26 & 2.26 & 0.05 \\
G037-037 & 105 & 5990 & 3.76 & $-$2.35 & 34.5$\pm$1.4  & 2.28 & 2.28 & 0.07 \\
G005-035 & 120 & 5439 & 4.18 & $-$1.76 & 26.0$\pm$1.2  & 1.71 & 1.77 & 0.05 \\
G246-038 & 680 & 5057 & 4.24 & $-$2.32 & $<$0.9$\pm$0.2 & $<-$0.16 &\nodata &\nodata \\
G095-060 &  90 & 5169 & 4.51 & $-$1.64 & 10.0$\pm$1.7  & 0.99 & 1.09 & 0.05 \\
G082-005 & 200 & 5378 & 4.43 & $-$0.71 & 7.0$\pm$0.7   & 0.90 & 0.98 & 0.04 \\
G082-023 & 135 & 4934 & 3.64 & $-$3.49 & $<$4.4$\pm$1.1 & $<$0.44 & \nodata &\nodata \\
G247-027 & 160 & 5016 & 4.40 & $-$1.97 & $<$5.9$\pm$0.9 & $<$0.61 & \nodata &\nodata \\
G084-052 &  70 & 5039 & 4.83 & $-$1.64 & $<$8.4$\pm$2.1 & $<$0.74 &\nodata &\nodata  \\
G097-040 &  80 & 5427 & 4.62 & $-$1.52 & 9.0$\pm$1.9   & 1.20 & 1.27: & 0.05 \\
LTT 2415 & 140 & 6295 & 4.11 & $-$2.13 & 23.5$\pm$1.1  & 2.31 & 2.28 & 0.06 \\
G110-034 & 100 & 5686 & 4.11 & $-$2.11 & 35.0$\pm$1.5  & 2.07 & 2.10 & 0.06 \\
G088-032 & 120 & 6136 & 3.54 & $-$2.55 & 22.0$\pm$1.2  & 2.17 & 2.17 & 0.06 \\
G088-042 &  90 & 5192 & 4.20 & $-$1.78 & 11.5$\pm$1.7  & 1.08 & 1.19 & 0.05 \\
G090-036 & 100 & 5319 & 4.14 & $-$1.76 & 25.5$\pm$1.5  & 1.59 & 1.67 & 0.06 \\
G251-024 &  90 & 5569 & 3.47 & $-$1.76 & 21.0$\pm$1.7  & 1.73 & 1.79 & 0.09 \\
G046-005 & 150 & 5191 & 4.83 & $-$1.41 & $<$3.9$\pm$1.0 & $<$0.53 &\nodata &\nodata \\
G009-036 & 140 & 5788 & 4.35 & $-$1.12 & 34.0$\pm$1.1  & 2.15 & 2.17 & 0.06 \\
G114-042 & 110 & 5761 & 4.34 & $-$1.11 & 22.5$\pm$1.3  & 1.92 & 1.95 & 0.09 \\
G116-053 & 100 & 5732 & 4.45 & $-$1.03 & 20.0$\pm$1.5  & 1.84 & 1.87 & 0.06 \\
G121-012 & 185 & 6163 & 4.37 & $-$0.80 & 45.0$\pm$0.8  & 2.59 & 2.55 & 0.05 \\
G197-030 &  95 & 5111 & 4.94 & $-$1.74 & $<$6.2$\pm$1.6 & $<$0.72 &$<$0.82: &\nodata \\
G122-051 & 700 & 5102 & 4.67 & $-$1.34 & 3.5$\pm$0.2  & 0.35 & 0.44: & 0.04 \\
G011-044 & 120 & 5924 & 3.82 & $-$2.23 & 33.0$\pm$1.2  & 2.21 & 2.28 & 0.06 \\
G238-030 & 210 & 5383 & 3.43 & $-$3.60 & 25.0$\pm$0.7  & 1.62 & 1.70: & 0.05 \\
G064-012 & 310 & 6074 & 3.72 & $-$3.45 & 24.0$\pm$0.5  & 2.15 & 2.15: & 0.06 \\
G165-039 & 210 & 6118 & 3.53 & $-$2.18 & 23.5$\pm$0.7  & 2.20 & 2.19 & 0.05 \\
G064-037 & 295 & 6122 & 3.87 & $-$3.28 & 15.5$\pm$0.5  & 1.97 & 1.97: & 0.05 \\
G166-037 & 110 & 5350 & 4.71 & $-$1.40 & 11.0$\pm$1.3  & 1.20 & 1.28: & 0.07 \\
G201-005 & 205 & 6018 & 3.79 & $-$2.50 & 27.0$\pm$0.7  & 2.17 & 2.17 & 0.06 \\
G239-026 & 160 & 5862 & 4.31 & $-$2.20 & 41.0$\pm$0.9  & 2.28 & 2.28 & 0.05 \\
G015-013 & 200 & 5082 & 4.61 & $-$1.72 & 3.0$\pm$0.7  & 0.33 & 0.41: & 0.05 \\
G016-025 & 120 & 5431 & 4.28 & $-$1.85 & $<$4.9$\pm$1.2 & $<$0.94 & $<$1.03&\nodata \\ 
G180-024 & 240 & 6108 & 4.12 & $-$1.40 & 32.0$\pm$0.6  & 2.35 & 2.33 & 0.08 \\
G168-042 & 185 & 5486 & 4.80 & $-$0.84 & 9.0$\pm$0.8   & 1.18 & 1.25: & 0.07 \\
G170-021 & 100 & 5664 & 4.65 & $-$1.47 & 24.0$\pm$1.5  & 1.88 & 1.91: & 0.07 \\
G020-008\tablenotemark{a} & 215 & 5960 & 4.04 & $-$2.33 & 36.0$\pm$0.7 & 2.28 & 2.28 & 0.05 \\
G020-008\tablenotemark{b} & 150 & 5940 & 3.91 & $-$2.34 & 33.0$\pm$1.0 & 2.22 & 2.23 & 0.05 \\
G184-007 & 115 & 5147 & 4.85 & $-$1.63 & $<$5.1$\pm$1.3 & $<$0.64 & $<0.73:$&\nodata\\ 
G262-021 & 125 & 4985 & 4.26 & $-$1.44 & $<$4.7$\pm$1.2 & $<$0.37 & \nodata &\nodata \\
G144-028 & 140 & 5310 & 4.19 & $-$2.14 & 21.5$\pm$1.1  & 1.51 & 1.59 & 0.05 \\
G025-024 & 180 & 5733 & 3.98 & $-$1.96 & 40.5$\pm$0.8  & 2.19 & 2.21 & 0.05 \\
G093-001 &  65 & 5430 & 4.37 & $-$1.59 & 17.5$\pm$2.3  & 1.51 & 1.58 & 0.08 \\
G026-012 & 320 & 6089 & 4.04 & $-$2.49 & 28.0$\pm$0.5  & 2.24 & 2.23 & 0.05 \\
G188-020 & 170 & 6174 & 4.14 & $-$1.58 & 35.5$\pm$0.9  & 2.45 & 2.42 & 0.05 \\
G188-030 & 350 & 5141 & 4.44 & $-$1.90 & 7.0$\pm$0.4   & 0.82 & 0.91 & 0.04 \\
G018-040 & 140 & 5681 & 4.18 & $-$1.93 & 36.5$\pm$1.1  & 2.09 & 2.11 & 0.05 \\
G241-004 & 110 & 5139 & 5.00 & $-$1.57 & $<$5.4$\pm$1.3 & $<$0.67 & $<$0.77:&\nodata \\
G215-047 & 150 & 5854 & 4.25 & $-$1.39 & 38.0$\pm$1.0  & 2.25 & 2.26 & 0.07 \\
G233-026 &  80 & 5303 & 4.39 & $-$1.45 & 9.0$\pm$1.9   & 1.06 & 1.15 & 0.05 \\
G189-050 & 180 & 5254 & 4.32 & $-$1.47 & 23.5$\pm$0.8  & 1.46 & 1.55 & 0.06 \\
G242-019 & 155 & 5039 & 4.20 & $-$2.12 & $<$3.8$\pm$1.0 & $<$0.45 & \nodata &\nodata \\
\enddata
\tablenotetext{a}{KPNO data}
\tablenotetext{b}{KECK data}
\end{deluxetable}
%\end{document}

\clearpage
%\documentclass[12pt,preprint]{aastex}
%\documentstyle[apjpt4]{article}

%\begin{document}
\begin{deluxetable} {ccccccccccccccccccc}
\tabletypesize{\scriptsize}
%\rotate
\tablewidth{0pt}
\tablenum{2}
\tablecolumns{19}
\tablecaption{Element Abundances in Plateau Stars}
\tablehead{
\multicolumn{1}{c}{Star}    &
\multicolumn{1}{c}{$T_{\rm eff}$}    &
\multicolumn{1}{c}{log g}    &
\multicolumn{1}{c}{[Fe/H]}    &
\multicolumn{1}{c}{$\sigma$[Fe/H]}  & 
\multicolumn{1}{r}{A(Li)}   &
\multicolumn{1}{r}{$\sigma$A(Li)}   &
\multicolumn{1}{c}{[Mg/H]}    &
\multicolumn{1}{c}{$\sigma$[Mg/H]}  & 
\multicolumn{1}{c}{[Ca/H]}    &
\multicolumn{1}{c}{$\sigma$[Ca/H]}  & 
\multicolumn{1}{c}{[Ti/H]}    &
\multicolumn{1}{c}{$\sigma$[Ti/H]}  & 
\multicolumn{1}{c}{[Cr/H]}    &
\multicolumn{1}{c}{$\sigma$[Cr/H]}  & 
\multicolumn{1}{c}{[Ni/H]}    &
\multicolumn{1}{c}{$\sigma$[Ni/H]}  & 
\multicolumn{1}{c}{[Ba/H]}    &
\multicolumn{1}{c}{$\sigma$[Ba/H]}  \\
}
\startdata
%STAR     & Teff & grav & [Fe/H]  & sigFe & A(LI) & sig  & [Mg/H]  &  sig & [Ca/H]  & sig  &  [Ti/H] &  sig &  [Cr/H] &  sig &  [Ni/H] &  sig &  [Ba/H] &  sig 
G171-050  & 5896 & 4.03 & $-$2.74 & 0.06  & 2.11  & 0.05 & $-$2.31 & 0.16 & $-$2.39 & 0.15 & $-$2.43 & 0.09 & $-$2.90 & 0.06 & $-$2.68 & 0.09 & $-$2.52 & 0.06 \\
G037-037  & 5990 & 3.76 & $-$2.35 & 0.06  & 2.28  & 0.07 & $-$2.14 & 0.29 & $-$2.01 & 0.09 & $-$2.01 & 0.08 & $-$2.42 & 0.17 & $-$2.42 & 0.09 & $-$2.39 & 0.09 \\ 
LTT 2415  & 6295 & 4.11 & $-$2.13 & 0.05  & 2.31  & 0.06 & $-$1.88 & 0.16 & $-$1.78 & 0.07 & $-$1.71 & 0.09 & $-$2.17 & 0.07 & $-$2.16 & 0.10 & $-$2.14 & 0.07 \\
G088-032  & 6136 & 3.54 & $-$2.55 & 0.06  & 2.17  & 0.06 & $-$2.02 & 0.16 & $-$2.18 & 0.06 & $-$2.09 & 0.09 & $-$2.54 & 0.14 & $-$2.45 & 0.16 & $-$2.96 & 0.13 \\
G011-044  & 5924 & 3.82 & $-$2.23 & 0.06  & 2.21  & 0.06 & $-$1.83 & 0.10 & $-$1.82 & 0.12 & $-$1.90 & 0.08 & $-$2.23 & 0.08 & $-$2.17 & 0.16 & $-$2.73 & 0.08 \\
G064-012  & 6074 & 3.72 & $-$3.45 & 0.04  & 2.15  & 0.06 & $-$2.82 & 0.16 & $-$2.84 & 0.09 & $-$2.95 & 0.11 & $-$3.38 & 0.28 & $-$3.73 & 0.16 & $-$3.65 & 0.13 \\
G165-039  & 6118 & 3.53 & $-$2.18 & 0.04  & 2.20  & 0.05 & $-$1.91 & 0.25 & $-$1.69 & 0.10 & $-$1.74 & 0.05 & $-$2.14 & 0.09 & $-$2.16 & 0.09 & $-$2.16 & 0.09 \\
G064-037  & 6122 & 3.87 & $-$3.28 & 0.04  & 1.97  & 0.05 & $-$2.81 & 0.16 & $-$2.84 & 0.10 & $-$2.77 & 0.06 & $-$3.50 & 0.15 & $-$3.38 & 0.16 & $-$3.77 & 0.16 \\
G201-005  & 6018 & 3.79 & $-$2.50 & 0.05  & 2.17  & 0.06 & $-$2.08 & 0.16 & $-$2.16 & 0.09 & $-$2.14 & 0.06 & $-$2.44 & 0.14 & $-$2.41 & 0.07 & $-$3.01 & 0.09 \\
G239-026  & 5862 & 4.31 & $-$2.20 & 0.05  & 2.28  & 0.05 & $-$1.94 & 0.06 & $-$1.97 & 0.10 & $-$2.02 & 0.08 & $-$2.30 & 0.09 & $-$2.37 & 0.22 & $-$2.39 & 0.05 \\
G020-008  & 5960 & 4.04 & $-$2.33 & 0.06  & 2.28  & 0.05 & $-$1.89 & 0.09 & $-$1.91 & 0.09 & $-$1.93 & 0.10 & $-$2.34 & 0.08 & $-$2.33 & 0.06 & $-$2.63 & 0.06 \\ 
G025-024  & 5733 & 3.98 & $-$1.96 & 0.05  & 2.19  & 0.05 & $-$1.67 & 0.06 & $-$1.66 & 0.07 & $-$1.75 & 0.06 & $-$2.01 & 0.07 & $-$1.99 & 0.10 & $-$1.88 & 0.14 \\
G026-012  & 6089 & 4.04 & $-$2.49 & 0.05  & 2.24  & 0.05 & $-$2.08 & 0.08 & $-$2.05 & 0.11 & $-$2.09 & 0.07 & $-$2.53 & 0.08 & $-$2.59 & 0.14 & $-$2.94 & 0.06 \\
G188-020  & 6174 & 4.14 & $-$1.58 & 0.05  & 2.45  & 0.05 & $-$1.26 & 0.05 & $-$1.22 & 0.09 & $-$1.22 & 0.06 & $-$1.57 & 0.09 & $-$1.54 & 0.17 & $-$1.39 & 0.17 \\
\enddata
\end{deluxetable}
%\end{document}

\clearpage
%\documentclass[12pt,preprint]{aastex}
%\documentstyle[apjpt4]{article}

%\begin{document}

\begin{deluxetable} {cccrr}
\tablecolumns{5}
\tablewidth{0pc}
\tablenum{3}
\tablecaption{Data for Stars in Figures 10 and 11}
\tablehead{
\multicolumn{1}{c}{Star}    &
\multicolumn{1}{c}{$T_{\rm eff}$}    &
\multicolumn{1}{c}{[Fe/H]}    &
\multicolumn{1}{r}{A(Li)}   &
\multicolumn{1}{r}{A(Li)$_{NLTE}$}   
}
\startdata
%STAR     & Teff & [Fe/H]  & A(Li) & A(LI)NLTE 
\cutinhead{[Fe/H] = $-$0.70 to $-$1.03;  $<$[Fe/H]$>$ = $-$0.86 }
G082-005  & 5378  & $-$0.71  & 0.90 & 0.98 \\
G168-042  & 5486  & $-$0.84  & 1.18 & 1.25 \\
G116-053  & 5732  & $-$1.03  & 1.84 & 1.87 \\
\cutinhead{[Fe/H] = $-$1.24 to $-$1.47;  $<$[Fe/H]$>$ = $-$1.41 }
G031-026 & 5546  & $-$1.32   & 1.59 & 1.64 \\ 
G122-051 & 5102  & $-$1.34   & 0.35 & 0.44 \\
G166-037 & 5350  & $-$1.40   & 1.20 & 1.28 \\
G170-021 & 5664  & $-$1.47   & 1.88 & 1.91 \\
G233-026 & 5303  & $-$1.45   & 1.06 & 1.15 \\
G189-050 & 5254  & $-$1.47   & 1.46 & 1.55 \\
\cutinhead{[Fe/H] = $-$1.58 to $-$1.90;  $<$[Fe/H]$>$ = $-$1.74 }
G005-035  & 5439  & $-$1.76   & 1.71 & 1.77 \\ 
G095-060  & 5169  & $-$1.64   & 0.99 & 1.09 \\
G088-042  & 5192  & $-$1.78   & 1.08 & 1.19 \\
G090-036  & 5319  & $-$1.76   & 1.59 & 1.67 \\
G015-013  & 5082  & $-$1.72   & 0.33 & 0.41 \\
G093-001  & 5430  & $-$1.59   & 1.51 & 1.58 \\
G188-030  & 5141  & $-$1.90   & 0.82 & 0.91 \\
\enddata
\end{deluxetable}
%\end{document}

\end{document}